\documentclass[format=sigconf, preprint]{acmart}
\pdfoutput=1

\renewcommand\footnotetextcopyrightpermission[1]{} 
 \usepackage{hyperref}

\usepackage{array}
\usepackage{algorithm}
\usepackage{algorithmicx}
\usepackage[noend]{algpseudocode}
\usepackage{algpseudocode}
\usepackage{epstopdf}
 
\usepackage{amssymb}  
\usepackage{amsmath}

\usepackage[T1]{fontenc} 
\usepackage[utf8]{inputenc}
\DeclareRobustCommand{\stirling}{\genfrac\{\}{0pt}{}}
\usepackage[normalem]{ulem}

\algrenewcommand\algorithmicindent{0.8em}%
\algrenewcommand{\algorithmiccomment}[1]{\hskip1em// #1}
\algrenewcommand{\algorithmicrequire}{\textbf{Input:}}
\algrenewcommand{\algorithmicensure}{\textbf{Output:}}
\algrenewcommand\textproc{}

\newcommand{\refequ}[1]{Equation~(\ref{equ:#1})}
\newcommand{\refsec}[1]{Section~\ref{sec:#1}}

\newcommand{\reffig}[1]{Figure~\ref{fig:#1}}

\newcommand{\reftab}[1]{Table~\ref{tab:#1}}
\newcommand{\refalg}[1]{Algorithm~\ref{alg:#1}}
\newcommand{\reflin}[1]{Line~\ref{lin:#1}}

\newcommand{\refequto}[2]{Equations~(\ref{equ:#1})--(\ref{equ:#2})}

\newcommand{\reflinto}[2]{Lines~\ref{lin:#1}--\ref{lin:#2}}

\newcommand{\fo}{FO}
\newcommand{\afo}{AFO}

\newcommand{\flo}{FO}

\newcommand{\ma}{HMAMA}

\newtheoremstyle{mydefinition}
  {.15em} 
  {.15em}  
  {} 
  {} 
  {\bfseries} 
  {.} 
  {.5em} 
  {} 

\theoremstyle{mydefinition}
 \renewcommand\footnotetextcopyrightpermission[1]{} 

\begin{document}
 
\title{Day-ahead Trading of Aggregated Energy Flexibility - Full Version }

\author{Emmanouil~Valsomatzis}
\affiliation{%
  \institution{Aalborg University}
  \country{Denmark}}
\email{evalsoma@cs.aau.dk}
\author{Torben Bach Pedersen}
\affiliation{%
  \institution{Aalborg University}
  \country{Denmark}}
\email{tbp@cs.aau.dk}
\author{Alberto  Abell\'o}
\affiliation{%
  \institution{Universitat Polit\`ecnica de Catalunya}
  \country{Spain}}
\email{aabello@essi.upc.edu}

\begin{abstract}
Flexibility of small loads, in particular from Electric Vehicles (EVs),  
has recently attracted a lot of interest
due to their possibility of participating in the 
energy market and the new commercial potentials.
Different from existing work, the aggregation techniques proposed in this paper produce flexible  aggregated loads from EVs taking into account technical market requirements. 
They can be further transformed into the so-called  flexible orders and be traded in the day-ahead market by a Balance Responsible Party (BRP). 
As a result, the BRP can achieve at least $20$\% cost reduction on average in   energy purchase compared to traditional charging    based on 2017 real electricity prices from the Danish electricity market.  
\end{abstract}

%
%



\maketitle
\footnotetext{Short version to appear as Note paper in the Proceedings of the e-Energy 2018, ninth ACM International Conference on Future Energy Systems (ACM e-Energy 2018) }
\section{Introduction}
\label{sec:intro}
The integration of EVs into the Smart Grid   reveals new business opportunities by exploiting their inherent flexibility~\cite{Hu20161207,Kristoffersen20111940}.
A market actor that     
controls the charging rate and time of a   portfolio of EVs   could acquire financial gain from energy arbitrage~\cite{5978239, Biegel2014354}. 
The energy required to charge (and/or discharge) the EVs can be traded through bids in  day-ahead and/or regulation market at a minimum cost~\cite{7328775}. 
Numerous   research studies, European (e.g.,~\cite{mirabelonline}, and national projects (e.g.,~\cite{totalflexonline})  focus on     trading   the required energy to charge EVs   taking into account different parameters.

An optimization charging approach of EVs that activates the participation in both day-ahead and regulation markets is proposed in~\cite{6003808}.
Scheduling techniques of EV charging that aim to maximize  the market actor's profit and take into account electricity price uncertainty are suggested in~\cite{7056534} and~\cite{7346507}. 
A risk-based scheduling framework for charging EVs is also proposed in~\cite{6740918}.
The suggested algorithm is based on day-ahead prices and takes into account driving activity uncertainties in order to minimize the charging cost of the EVs. 
Similarly, a day-ahead optimization technique for scheduling EVs considering the impact  on the day-ahead prices is suggested in~\cite{7879192}.
Both optimization and heuristic techniques for optimal charging of EVs aiming to the maximization of the revenue by utilizing energy storage are proposed in~\cite{6461500}.

The main characteristic of the research  tackling the energy trading of flexible EV loads is the output of the proposed techniques, i.e., an aggregated scheduled load.
Unlike other work, we introduce $3$ aggregation techniques that  
produce {\em flexible} aggregated loads that can be traded in the market.
As a result, the market itself,   not the market actors, schedules the loads from the EVs as part of the trading process,  minimizing the uncertainty of bidding.
For instance, instead of placing a bid to purchase $30$MW in hour $3$, 
the market actor places a bid to purchase $30$MW in any hour between hour $1$ and $5$.
The market  determines the activation time of the bid. 
In many cases, the technical trading details of the market impose hard constraints that are omitted by the proposed solutions and the realization of the suggested techniques becomes very difficult.
For instance, the proposed scheduling technique in~\cite{7210197} offers less than $200$kW in less than an hour in the regulation power market where the minimum bid is $10$MW, in full hours~\cite{Biegel2014354}.
The bidding strategies proposed in~\cite{Baringo2017362}  and the high power fluctuations of the scheduling 
outputs in~\cite{6003808} and in~\cite{7328775}, would require single hour independent bids~\cite{Nordpool} that might not fulfill the energy requirements.
The conservative bidding approach for the bidding strategy proposed in~\cite{6935028} covers less than 50\% of the energy needed to charge the EVs.
On the contrary, our proposed   aggregation techniques  use real technical market requirements derived from a specific order (bid) type of Nordic market, namely, {\em flexible orders}~\cite{Nordpool}.

\textbf{Contributions.}
First, we describe both the so-called flex-offer (\fo{}) model, which captures the flexibility of the EVs, and a realistic market framework where the flexibility is traded.
Second, we investigate the market-based FO aggregation problem and its complexity. 
Third, we introduce $3$ heuristic algorithms that take into account real market requirements and produce flexible aggregated FOs that can be then traded through flexible orders in the market.
Finally, we compare our proposed techniques with $2$ base-line approaches and
evaluate both the technical and the financial aspect of   their results based on real market prices.
We show that our proposed techniques achieve more than $20$\% cost reduction on average in the purchased energy required to  charge  from $5$K to $40$K EVs.

The paper is organized as follows:
we introduce the preliminary definitions in~\refsec{preliminaries} and 
we present the problem formulation of market-based aggregation in~\refsec{problemFormulation}.
In~\refsec{alg}, we propose $3$ heuristic market-based aggregation techniques and we experimentally evaluate them in~\refsec{exp}. 
We conclude the paper in~\refsec{con}.  
\section{Preliminaries}
\label{sec:preliminaries}
In this section, we describe the EV   model that can be used to trade flexibility and the market framework     used for trading.

\subsection{Electric vehicle model} 
\label{sec:EVmodel}
We consider the energy used to charge  EVs to be appropriate for flexible energy trading. 
The reason is that   the lithium-ion batteries of EVs are ample power demand devices and their charge can be time shifted when the EVs are plugged-in for  more hours than needed for charging. 
We consider EVs that can   be  continuously charged with a   power-constant voltage (CP-CV) option~\cite{7098444}  and their charge is taking place in the range of   20\% to 90\%
state of charge (SOC)  so that      the battery life is preserved~\cite{6345063}.
As a result,   when an EV     is plugged in for charging, its battery capacity is at least 20\% and the user would like to fully charge it (90\%) for his/her next trip.
The SOC is computed according to the following formula based on~\cite{7098444}:
\begin{math}
\mathit{SOC_{final}} = \mathit{SOC_{ini}} + \frac{\eta_c \cdot \eta_b\cdot P \cdot time_{cha}}{C}
\end{math} (1)
 where \begin{math}\mathit{SOC_{final}}\end{math} is the final state of charge equal to   90\% of the total battery capacity ($C$).
 Parameters  \begin{math} \eta_c\end{math} and \begin{math}\eta_b\end{math}   represent    the efficiency of the charger 
 and the internal resistance of the battery, respectively.
We  represent with $P$ the power used to charge an EV 
that is constant over the [20\%, 90\%] interval of SOC and $ time_{cha}$ is the time needed to   charge it up to $\mathit{SOC_{final}}$.
 

In our work, because we take into account  time shifted loads, we use   the flex-offer (FO) concept~\cite{TKDEEManolis}, introduced in the MIRABEL project~\cite{mirabelonline,pub:mirabel_endm2012} 
to represent the charging of a flexible EV.
An FO captures flexibility from different dimensions (e.g., time, energy, and/or combined)~\cite{Valsomatzis15},
from different devices~\cite{Siksnys:2016:DFS:2934328.2934339}, and
can be used for different purposes, e.g., tackle electrical grid bottlenecks~\cite{7778808}.
Thus, we define an {\em \flo{}   $f$ }to be a tuple \begin{math}f=(T(f), P(f))\end{math} where $T(f)$ is the start charging flexibility interval and 
$P(f)$ is the power profile. 
\begin{math}T(f)=[t_{es}, t_{ls}]\end{math}  
where $t_{es}$ and $t_{ls}$ are the {\em earliest start charging time} and {\em latest start charging time}, respectively. 
We define time flexibility ($\mathit{tf}$) to be the difference between $t_{ls}$ and $t_{es}$.
The {\em power profile} is a sequence of (\begin{math}m\in\mathbb{N}_{>0}\end{math}) consecutive   slices, 
\begin{math}P(f)=\langle s^{(1)},\dots, s^{(m)}\rangle\end{math}   where a {\em slice} $s^{(i)}$ has a power value $p$ measured in kW. 
The duration of slices is 1 hour. 
 
For instance, 
an EV is plugged in at a house between $1$  and $8$ a.m.
The EV continuously utilizes $3.7$kW   for  $3.3$ hours to be charged. 
However, energy trading is performed per hour and we also use hourly resolution to model the EVs charging.
To respect the hourly granularity, we equally distribute the sum of the energy needed during the first and the last   regular charging hours and we reduce power fluctuations in the model. 
Therefore, we assume that the
EV  consumes $2.4$kWh both during the first and the last   charging hours and $3.7$kWh during the   hours in-between. 
The  EV   can be modeled by   an FO  
$f$$=$$([1,4],\langle2.4,  3.7, 3.7, 2.4\rangle)$, see~\reffig{FOandFO}a.
Next, we describe the market framework where such FOs shall be traded.
  
\subsection{Market framework}
\label{sec:marketFramework}
The Nordic/Baltic market for electrical energy named Nord Pool is considered in our work.
Nord Pool is one of the most mature energy markets~\cite{Modeling}
and Europe's leading power market~\cite{Nordpool}.
It consists of  the day-ahead (Elspot) and   intra-day markets.
We focus on Elspot because it has one of 
the largest turnovers in the Nordic system and it also supports  flexible energy trading~\cite{Biegel2014354}.
Trading in Elspot occurs daily     through orders (bids).
Each day   before 12 p.m., the balance responsible parties (BRPs) place their purchase and/or selling orders (bids) in  Elspot for the following day.
The orders specify the energy amount   a BRP desires to buy/sell and the price the BRP is willing to pay/be paid for the corresponding energy. 
Since 2016, 
Elspot supports 4  different order types: 
single hourly orders (price dependent or independent),
block orders,
exclusive groups, and
{\em flexible orders}~\cite{Nordpool}. 
We focus on  flexible orders that support flexibility trading.

When a BRP places a flexible order in Elspot, it  states the {\em name}, the {\em time interval}, the {\em price limit}, the {\em volume}, and the {\em duration} of the order.
The time unit is one hour and  volume is expressed in MW.
The duration expresses the number of hours during which the order can be activated over the interval \begin{math}[1, 23]\end{math}.
The time interval must exceed the duration by at least one hour and expresses the potential activation times of the order.
Volume is either positive, if the order is a purchase order or negative, if it is a sell order.
A BRP can place  $5$ flexible orders during a trading day.

{\em Hypothetically}, a BRP could purchase   the energy needed to charge the above mentioned flexible EV, represented by     FO $f$ 
through a flexible   order. 
%
The duration of a flexible order is mapped to the number of   slices of $f$, the volume to the power of the slices, and  the time interval to the time flexibility of $f$.
For instance,   a BRP   could   place  a flexible order named ``F1'', with duration $4$ hours and time interval from $1$ to $8$.
The volume of F1 is $0.0037$MW (in order to satisfy all the slices) and its price limit is 35 euros/MWh.
However, the energy needed to charge a single EV is (much) too small to be traded in Elspot.
In particular, the  minimum contract size and  the volume trade lot  for a flexible order are both $100$kW, while the power used by an EV is a few kW.
Moreover, when the duration of a flexible order is more than one hour, the volume needed for these hours shall be constant.
As a result, it is necessary to {\em aggregate} FOs to trade   the flexible loads of the EVs   through flexible   orders in  Elspot market.

\begin{figure}[tb]
\begin{tabular}{cc}
\hspace{-0.15in}
\includegraphics[width=0.24\textwidth]{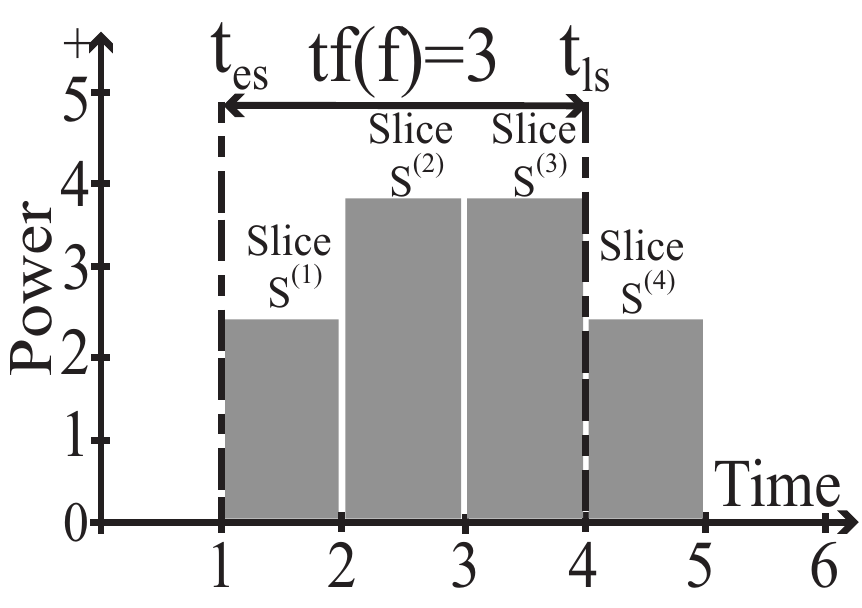}    &
\includegraphics[width=0.24\textwidth]{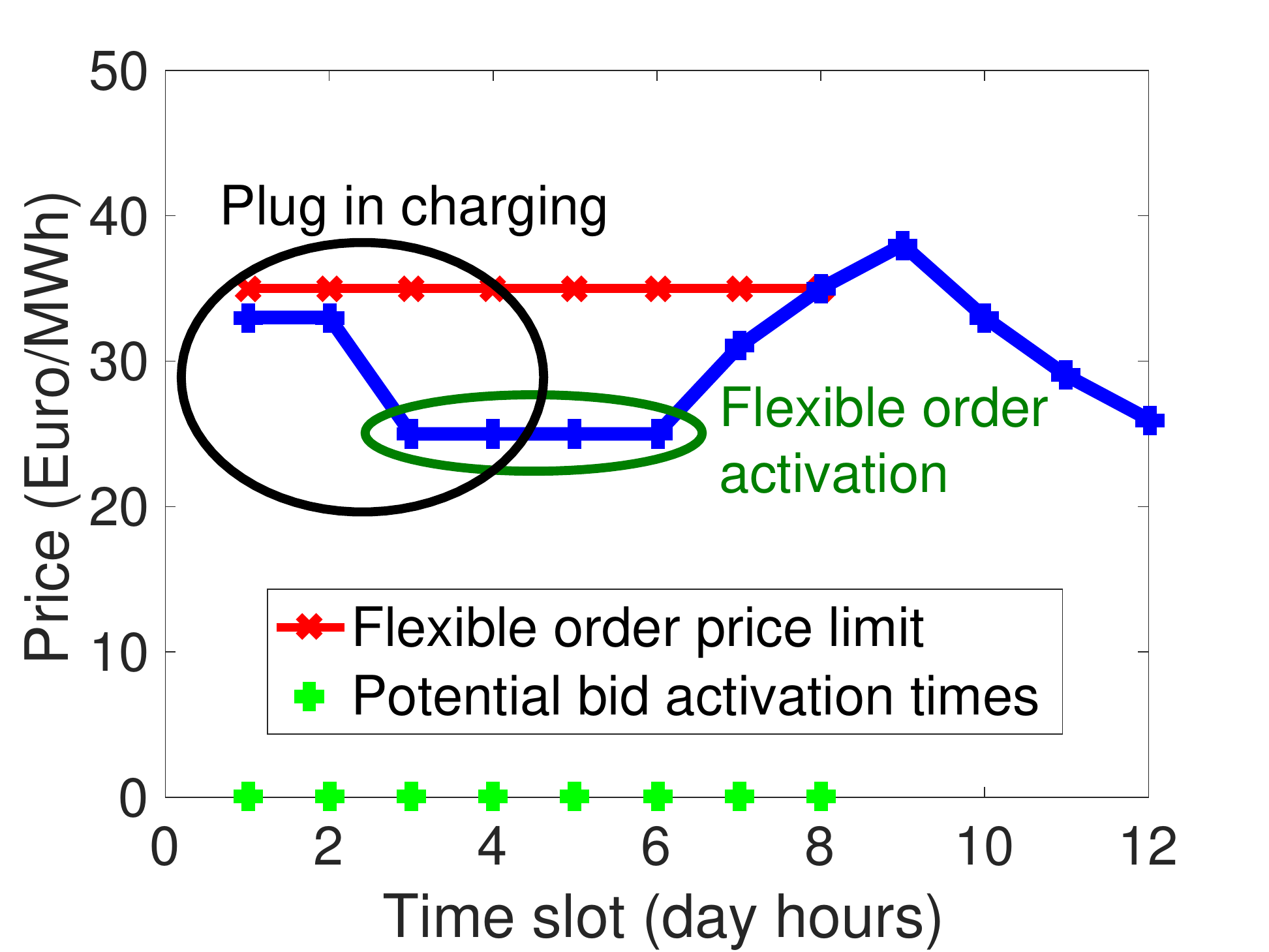}\\
(a) & (b) \\
\end{tabular}
\caption{An examle of an FO and a flexible purchase order\label{fig:FOandFO}} 
\end{figure}
 
The flexible 
order is activated in the time 
interval that optimizes social welfare provided that the price is respected~\cite{Nordpool}.
Given F1 in a liquid market, the order is activated when the cost of buying the required energy is minimized.
For instance, we see in~\reffig{FOandFO}b that F1  is activated in time slots   $3$, $4$, $5$, and $6$  where the price is $25$   euros/MWh.
Thus, the    energy needed to charge   the EV costs \begin{math}25\cdot0.0037\cdot4=0.37\end{math} euros.
On the contrary, 
if time flexibility  of the EV is disregarded, its charging occurs based on a price independent order and its plug-in time (time slot $1$-$4$ in~\reffig{FOandFO}b).
As a result,    the energy needed to charge the EV is purchased based on a price independent order and the price is set by Elspot.
In that case and according to~\reffig{FOandFO}b, the cost is \begin{math}33\cdot0.0037\cdot2 + 25\cdot0.0037\cdot2=0.4292\end{math} euros, $16$\% more than the cost achieved by   flexible order F1.
Therefore, a flexible order has a higher probability to achieve a better price than a price independent order because it takes into account the time flexibility of the flexible loads and thus  can be favored  by   price reductions.
The absolute difference (imbalance)  between the purchased energy and the energy needed  is traded in the balance market and usually for a higher price than the one in Elspot.
Consequently, the BRP desires to be as precise as possible regarding  the purchased energy from Elspot. 

Regarding the communication among an EV and a BRP, 
we assume an Information and Communication Technology (ICT) infrastructure~\cite{Albano2015133}.  
When an EV is plugged in, an FO is generated requiring the minimum  interaction with the owner of the EV.
The FO generation takes into account the historical use of the EV, the
$\mathit{SOC_{ini}}$ of the EV, the charging characteristics of the EV, and the technical  characteristics of the charging station, e.g., a home charging installation~\cite{Neupane:2017:GEF:3077839.3077850}.
Identifying the time flexibility of an FO is challenging, but appropriate forecast techniques can be designed   taking into account daily/weekly driving patterns~\cite{6465724}.
  

  
%
\section{Problem Formulation}
\label{sec:problemFormulation}
In this section, we show how aggregation of FOs that represent flexible charging loads of EVs can fulfill the requirements of flexible   orders. 
We also introduce the problem of market-based aggregation. 

\subsection{FO aggregation}
\label{sec:SA}
Based on~\cite{TKDEEManolis}, FO aggregation is the function that given a set of FOs $F$, produces a set of aggregated FOs   $\mathit{AF}$ where
\begin{math}\mathit{|AF|}\leq |F|\end{math}.
The produced AFOs capture large amounts of energy that can be traded in the market. 
Due to the time flexibility of the FOs, there are different alignment combinations that can lead to different AFOs.
According to start-alignment FO aggregation, 
the  earliest start charging time of an aggregated FO (AFO) $f_{a}$ is the minimum earliest start charging time
among all the FOs that produced it, i.e., 
\begin{math}f_{a}.t_{es} = min_{f\in F'}(f.t_{es}),F'\subseteq F\end{math}.
The    latest start charging time of $f_{a}$ is the sum of its $t_{es}$ and the minimum time flexibility among all the FOs in $F'$, i.e., \begin{math}f_{a}.t_{ls} = f_{a}.t_{es} + min_{f\in F'}(tf(f))\end{math}.
The power profile of  $f_{a}$ is produced by summing up the power profiles of the FOs when they are aligned according to  their earliest start charging time.

\begin{figure}[tb]
\centering
\includegraphics[width=0.47\textwidth]{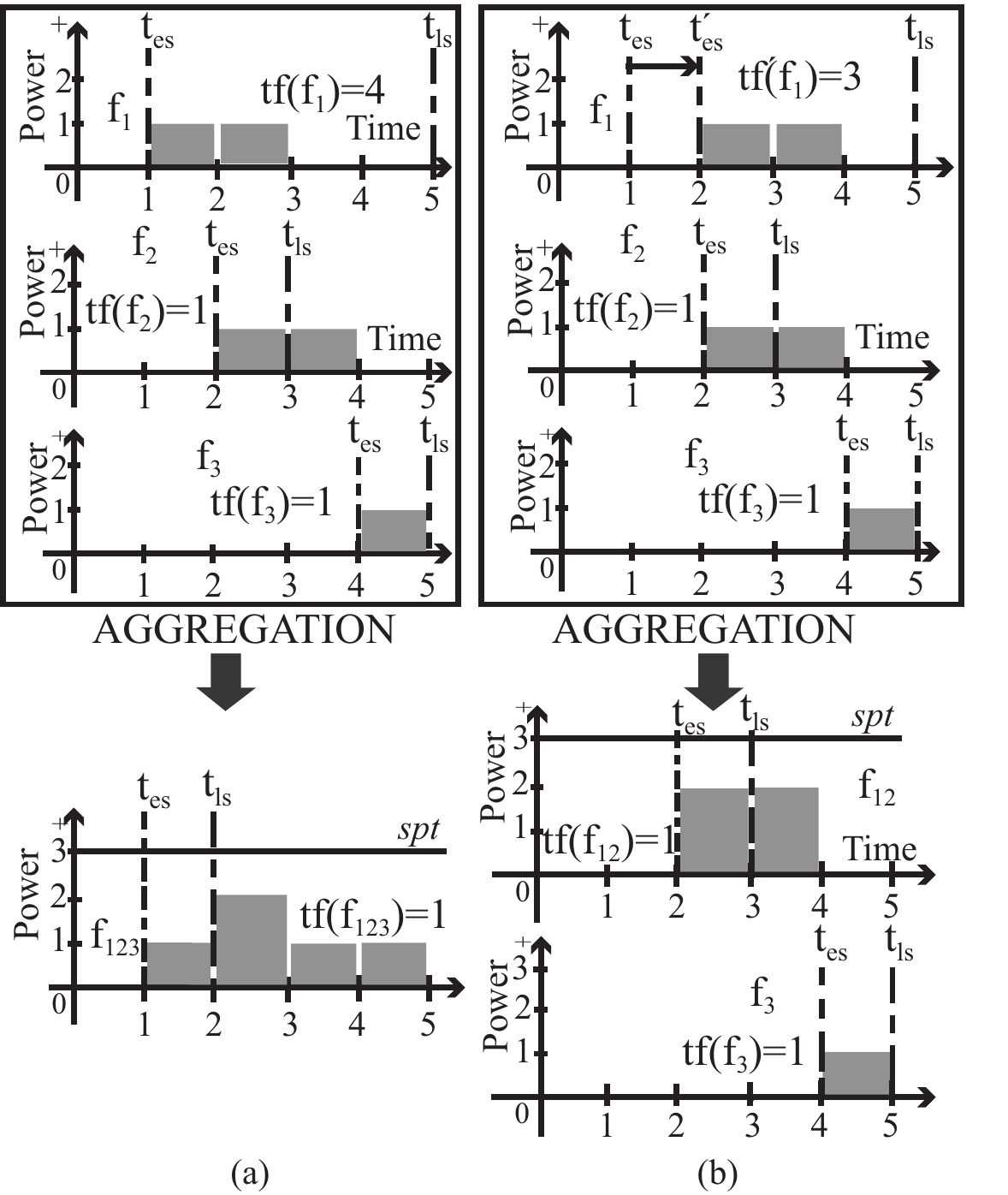}    
\caption{Flex-offer aggregation according to different alignments   \label{fig:alignments}} 
\end{figure}

For instance, we see in~\reffig{alignments}a three FOs, 
\begin{math}f_1=([1,5], \langle1,1\rangle)\end{math}
\begin{math}f_2=([2,3], \langle1,1\rangle)\end{math}, and
\begin{math}f_3=([4,5], \langle1\rangle)\end{math}, that produce   AFO $f_{123}$ 
where   \begin{math}f_{123}.t_{es}  = f_{1}.t_{es}=1\end{math} and
$f_{123}.t_{ls}$ is   the sum of $f_{123}.t_{es}$
and   time flexibility of $f_{2}$ or $f_{3}$, 
i.e., $f_{123}.t_{ls}=2$.
The power profile of  $f_{123}$ is produced by summing up the power profiles of $f_1$, $f_2$, and $f_3$  based on their alignments.
Thus,    
\begin{math}f_{123}.s^{(1)}.p = f_{1}.s^{(1)}.p =1\end{math},
\begin{math}f_{123}.s^{(2)}.p = f_{1}.s^{(2)}.p + f_{2}.s^{(1)}.p=2\end{math},  
\begin{math}f_{123}.s^{(3)}.p = f_{2}.s^{(2)}.p =1\end{math}, and
\begin{math}f_{123}.s^{(4)}.p = f_{3}.s^{(1)}.p =1\end{math}.

Due to the time flexibility of the FOs, there are different alignment combinations that can lead to different AFOs.
For instance, given the $3$ FOs \begin{math}f_1,f_2,f_3\end{math} in~\reffig{alignments} with time flexibility 4, 1, and 1, respectively, there are 20
\begin{math}(5\cdot2\cdot2)\end{math}  alignment combinations.
As a result, based on  different alignments,   time flexibility of the FOs can be adjusted accordingly and different power profiles for the AFOs are produced.
Moreover,   a set of FOs can be partitioned and each subset can produce an AFO.
Consequently, the output size of aggregation can be greater than one.
For instance, we see in~\reffig{alignments}b that the output of aggregation is $2$ AFOs, i.e., $f_{12}$ and $f_{3}$.
In particular, FO  $f_1$ is aligned with $f_2$ and time flexibility of $f_1$ is adjusted  so that   $f_1.t'_{es}$ is equal to  $f_2.t_{es}$.
Consequently, the power profiles of $f_1$ and $f_2$ are summed up and they produce AFO $f_{12}$.

\subsection{Market-based FO aggregation}
\label{sec:MBFO}

Given a portfolio, the goal of a BRP is to maximize its profit by purchasing, for the minimum price, the energy that it     sells to its customers.
We consider   flexible EVs to   be part of a BRP's portfolio and, since the  energy purchase takes place through orders, we examine if the energy needed to charge the flexible EVs can be purchased through flexible orders.
The purchasing strategy of a BRP depends on many   different factors, e.g., the content of the portfolio (factories, households, etc.) and   pricing forecast.
The   strategy is out of   scope of this work and left for Future work.
However, since a flexible order has in general a higher probability to achieve a lower purchase price, 
we consider the goal of a BRP to be the maximization of the purchased energy through flexible orders. 
In our work, we introduce {\em market-based FO aggregation} (MAGG) 
to be the aggregation that given a set of FOs, outputs between one and five AFOs  that fulfill the flexible order requirements, see~\refequto{eq1}{eq2}.
AFOs summarize the energy requirements and flexibilities in amount and time imposed by the 
technical requirements of a flexible order.

In order for an FO to fulfill the flexible order requirements, the FO must have ($1$) time flexibility at least one (\refequ{eq3}) and ($2$) between $1$ and $23$ slices (\refequ{eq4}). 
Moreover,   since the minimum contract size and the trade lot of a flexible order are both $100$kW, ($3$) the values of the slices of the FOs shall be multiples  of $100$kW.
 
For illustrating purposes, we assume in our example below that {\em both the   volume and the trade lot   for a flexible order is $2$kW instead of $100$kW}.
For instance, we see in~\reffig{alignments} that none of the individual   FOs fulfills the power profile requirements  of a flexible order ($2$kW).
Thus, market-based FO aggregation is necessary.
In that case, market-based FO aggregation  produces 
AFO $f_{12}$ that fulfills the flexible-order requirements 
since its time flexibility is 1  and the  power of both the slices  equals to 2, see~\reffig{alignments}b.
FO  $f_3$ is also part of the aggregation output, but it is not a valid AFO 
because it does not fulfill the power profile requirement, i.e., its slice amount is lower than 2.

The flexible EVs are represented by  a set of FOs. 
For instance, $5000$  EVs that are part of a BRP's portfolio   are represented by a set of FOs  $F$.
Each EV is an FO $f$ of the set, i.e., 
\begin{math} f \in F, f=(T(f),P(f)), T(f)=[t_{es}, t_{ls}], P(f) = \langle s^{(1)}, \dots, s^{(m)}\rangle\end{math}.
A BRP must aggregate the FOs to produce AFOs that  fulfill the flexible order requirements and can be then placed in the market as flexible orders. 
The volume of energy is expressed through the sum of the slices of the FOs and the power of each slice must be a multiple of 100kW (\refequ{eq6}).
However, due to technical charging characteristics (EV power demand is in the interval [3.7kW,11kW] for household charging),
we take into account a power range to define the valid power amounts.
Thus, instead of considering exact multiples of $100$kW for the power amount of each slice, we permit an insignificant amount  deviation of $e$kW per slice, e.g., $5$kW (\refequ{eq6}).
When the financial evaluation of market-based aggregation occurs, the  deviated amount  will be considered to be traded in balance market, see~\refsec{marketFramework}.
Hence, 
the problem of maximizing the bidden energy through flexible orders given a set of FOs is formulated as follows:
\begin{subequations}
\begin{align}
&{\text{Maximize}}
& &\sum_{f_a\in \mathit{AF}} \sum_{s\in P(f_a)} s.p\label{equ:eq1}\\
& \text{subject to}
& & \mathit{AF} = \mathit{MAGG(F)}, 1\leq|\mathit{AF}|\leq5  \label{equ:eq2}\\ 
& & & \forall f_a\in    \mathit{AF}, \mathit{tf(f_a)}\geq 1  \label{equ:eq3}\\ 
& & & \forall f_a\in    \mathit{AF}, 1\leq|P(f_a)|\leq 23 \label{equ:eq4}\\
& & &  \forall f_a  \in \mathit{AF}, \forall s\in P(f_a),  \label{equ:eq5}\\
& & &s.p=       x\cdot 100\text{kW} \pm e \text{kW},   x\in  \mathbb{N}_{> 0}, e\in [0,5] \label{equ:eq6} 
\end{align}
\end{subequations} 

\subsection{Market-based FO aggregation complexity.}

Given  a set of   \flo{}s $F$,   
there are $\stirling{|F|}{k}$ ways (Stirling numbers of the second kind~\cite{graham1994concrete}) to partition the $|F|$ \flo{}s   into $k$  
subsets. 
Applying aggregation on  each subset produces an AFO. 
In market-based FO aggregation, the size of the output  is between $1$ and $5$.
Thus, $k$ can be assigned values from $1$ to $5$.
Therefore, there are $\stirling{|F|}{1}$ ways to partition $|F|$ \flo{}s into $1$ non-empty subset of \flo{}s.
There are $\stirling{|F|}{2}$ ways to partition the $|F|$ \flo{}s    into $2$ non-empty subsets, where the aggregated \flo{}s are $2$  and so on.
Thus,     given  $|F|$ \flo{}s, there are 
\begin{math}\stirling{|F|}{1}+\stirling{|F|}{2}+\dots+\stirling{|F|}{5}=\sum_{k=1}^{5}\stirling{|F|}{k}\end{math} ways to partition the \flo{}s. 

Moreover, the number of the different aggregated \flo{}s depends on the alignments of the \flo{}s and thus on their time flexibility.   
In particular, given a set of \flo{}s $\mathit{SF}$ ($\mathit{SF}$$\subseteq$$F$) with time flexibility  \begin{math}\mathit{tf(f_1),\dots,tf(f_{|\mathit{SF}|})}\end{math} respectively, 
the number of the aggregation results (aggregated \flo{}s) that can be produced is:
\begin{math}\prod_{i=1}^{|\mathit{SF}|}\mathit{tf(f_i)}\end{math}.
Hence, given an average number of partitions (\begin{math}\mathit{avg(al)}\end{math}),
there are 
\begin{math}\sum_{k=1}^{5}\stirling{|F|}{k})\times\mathit{avg(al)}\end{math} potential aggregation results. 
Furthermore, 
the complexity of the  problem,    as an Integer Linear Programming problem,  is too high to be solved by state-of-the-art solvers~\cite{LPsolvers}. 


\begin{example}
Given a set with $100$ \flo{}s, there are \begin{math}\sum_{k=1}^{5}\stirling{100}{k}= 6.5738\cdot10^{67}\end{math} potential partitions that can produce from 1 to 5 AFOs.
Assuming $20$ alignments per partition on average,
there are in total 
\begin{math}20\cdot6.5738\cdot10^{67}=1.3148\cdot10^{69}\end{math} (approximately the estimated number of atoms in the Milky Way Galaxy) 
potential aggregation results that have to be examined in order to find the optimal one.
\end{example}
\section{Heuristic solutions}
\label{sec:alg}
Due to the unrealistically   large solution space, we   instead propose 3     variations of a  heuristic algorithm, i.e., {\em  Heuristic Market-based  Aggregation Main Algorithm} (\ma{})    that tackles the market-based aggregation problem. 
 
\subsection{Heuristic Market-based Aggregation  Main Algorithm}

The goal of \ma{} is to produce AFOs that respect the flexible order requirements while avoiding the high complexity of the problem and  at the same time provide good results in terms of bidden energy amount.
Thus, given a set of FOs $F$, \ma{}  (\refalg{ha}) performs incremental binary aggregations so that the produced AFOs increase the captured energy in each step. 
In addition, the algorithm maps the flexible order requirements to threshold parameters that must be respected during the performed aggregations.
Consequently, it introduces 3 thresholds, namely, the slice power ($\mathit{spt}$), time flexibility ($\mathit{tft}$), and power profile ($\mathit{ppt}$) thresholds that correspond to flexible order requirements. 
It sets  $\mathit{spt}$  to $100$   since flexible orders must have multiples of 100kW power.
Moreover, \ma{} assigns  $1$ and $23$ to $\mathit{tft}$     and   $\mathit{ppt}$, respectively, since flexible orders must have a time interval of $1$ and duration at most $23$ hours.
Permitted   amount  deviation is represented by $e$  that is assigned  values from $0$kW to $5$kW.

\begin{algorithm}[tb]
\begin{algorithmic}[1]
\Require{$F$ -  set  of \fo{}s, 
$e$ - amount deviation}
\Ensure{$\mathit{AF}$   -   set of A\fo{}s}
\State $continue\leftarrow true$, $\mathit{AF}\leftarrow \emptyset$
\While {$continue=true$}\label{lin:habodystart}
\State  $\mathit{ppt}\leftarrow23$, $\mathit{spt}\leftarrow100$   
\State{$\mathit{PF,UF},f_{ini},\mathit{tft}\leftarrow$Initialize($F$)} \label{lin:haini}
\State{$\mathit{PF,AF} \leftarrow$Process($\mathit{PF},\mathit{AF}, f_{ini},\mathit{tft},\mathit{ppt},\mathit{spt},e$)}  \label{lin:haprocess}
\State{$F, continue\leftarrow$Examine($\mathit{PF,UF},\mathit{AF},continue$)} \label{lin:haexamine}
\EndWhile\label{lin:habodyend}
\State\Return Top5EnergyAFOs($\mathit{AF})$\label{lin:hareturn}
\end{algorithmic}
\caption{Heuristic Market-Based Aggregation}\label{alg:ha}
\end{algorithm} 

\begin{algorithm}[tb]
\begin{algorithmic}[1]
\Function{Initialize}{$F$} 
\State $f_{ini}\leftarrow$SelectAmongLongestTheMostFlexibleFO($\mathit{F}$)\label{lin:lpini}   
\State\Return $ \mathit{F}\setminus f_{ini},\emptyset,f_{ini},1$\label{lin:returnini}
\EndFunction
\end{algorithmic}
\caption{Longest Profile - Initialization phase}\label{alg:ln}
\end{algorithm}

The body of \ma{} consists  of $3$ phases (functions), i.e.,  
{\em initialization},   {\em processing},  and  {\em examination} (\refalg{ha},~\reflinto{habodystart}{habodyend}).
During the initialization phase (\reflin{haini}), \ma{} identifies the FO with which to   start binary  aggregations ($f_{ini}$)   and the subset of the FOs  ($\mathit{PF}$) that   participates in the aggregations.
Then, during the processing phase (\reflin{haprocess}), it   produces all the potential binary aggregations between $f_{ini}$ and the FOs in $\mathit{PF}$ to produce AFOs that fulfill the flexible order requirements.
Afterwards, during the examination phase (\reflin{haexamine}), \ma{} examines whether it shall restart using the remaining FOs or terminate.

\subsection{Main Algorithm variants}

The initialization phase is salient for the outcome of the algorithm as it mainly defines the solution space that the algorithm explores.
Hence, we introduce $3$ variants of \ma{}   that have different initialization phases, namely, the {\em Largest Profile}  (LP), {\em Dynamic Profile} (DP), and {\em Dynamic Time Flexibility} (DTF).


LP focuses on producing AFOs with   many slices because a long FO  usually captures large energy amounts.
On the other hand, 
given an FO with many slices, it is very difficult to fulfill the flexible order amount requirements and, especially, the slice amount equality required.
For this reason, DP excludes from aggregation the FOs with extremely large profiles  (outliers).
DTF focuses on time flexibility of the FOs that  has a prominent role in aggregation since it is directly correlated to the alignments.   
Thus, DTF takes into account the time flexibility distribution of the initial set and  gradually   excludes     from aggregation the FOs with low time flexibility compared to the initial set.

\textbf{LP - Initialization phase.}
LP starts   by selecting the most 
flexible FO among the ones with the   largest profile size (\refalg{ln},~\reflin{lpini}).
An FO with large profile size and high time flexibility has high probability to time-wise overlap with profiles of other FOs.
So, AFOs that fulfill the flexible order requirements through different alignments can be produced. 
LP uses the initial set   $F$ as the  processing set $\mathit{PF}$ (\reflin{returnini}) 
and then executes the processing and examination phase. 
 
\begin{algorithm}[tb]
\begin{algorithmic}[1]
\Function{Initialize}{$F$}
\State $\mathit{uf}\leftarrow$UpperFenceProfileSize($F$)\label{lin:dpuf}   
\State $\mathit{PF}\leftarrow$FOsWithProfileAtLeastUF($\mathit{F,uf}$)\label{lin:dpal}  
\State $f_{ini}\leftarrow$SelectTheMostFlexibleFOAmongLongest($\mathit{PF}$)\label{lin:dpselectfnom}  
\State\Return $\mathit{PF}\setminus f_{ini}, F\setminus\mathit{PF}, f_{ini},1$\label{lin:dpreturn}
\EndFunction
 
\caption{Initialization phase - Dynamic Profile algorithm}\label{alg:dp}
\end{algorithmic}
\end{algorithm} 

\begin{figure}[tb]
\begin{tabular}{ c  c  }
\includegraphics[width=0.145\textwidth]{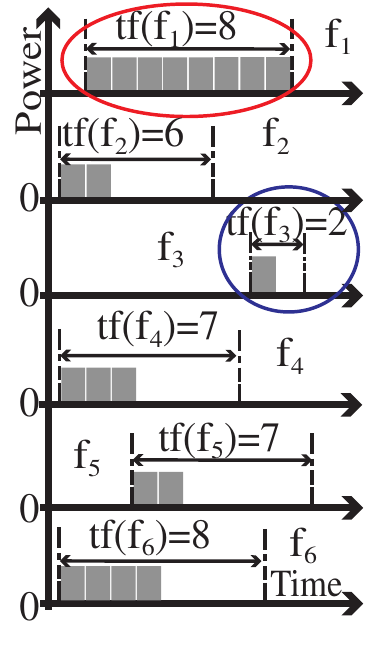}&
\hspace{-1.5em}
\includegraphics[width=0.35\textwidth]{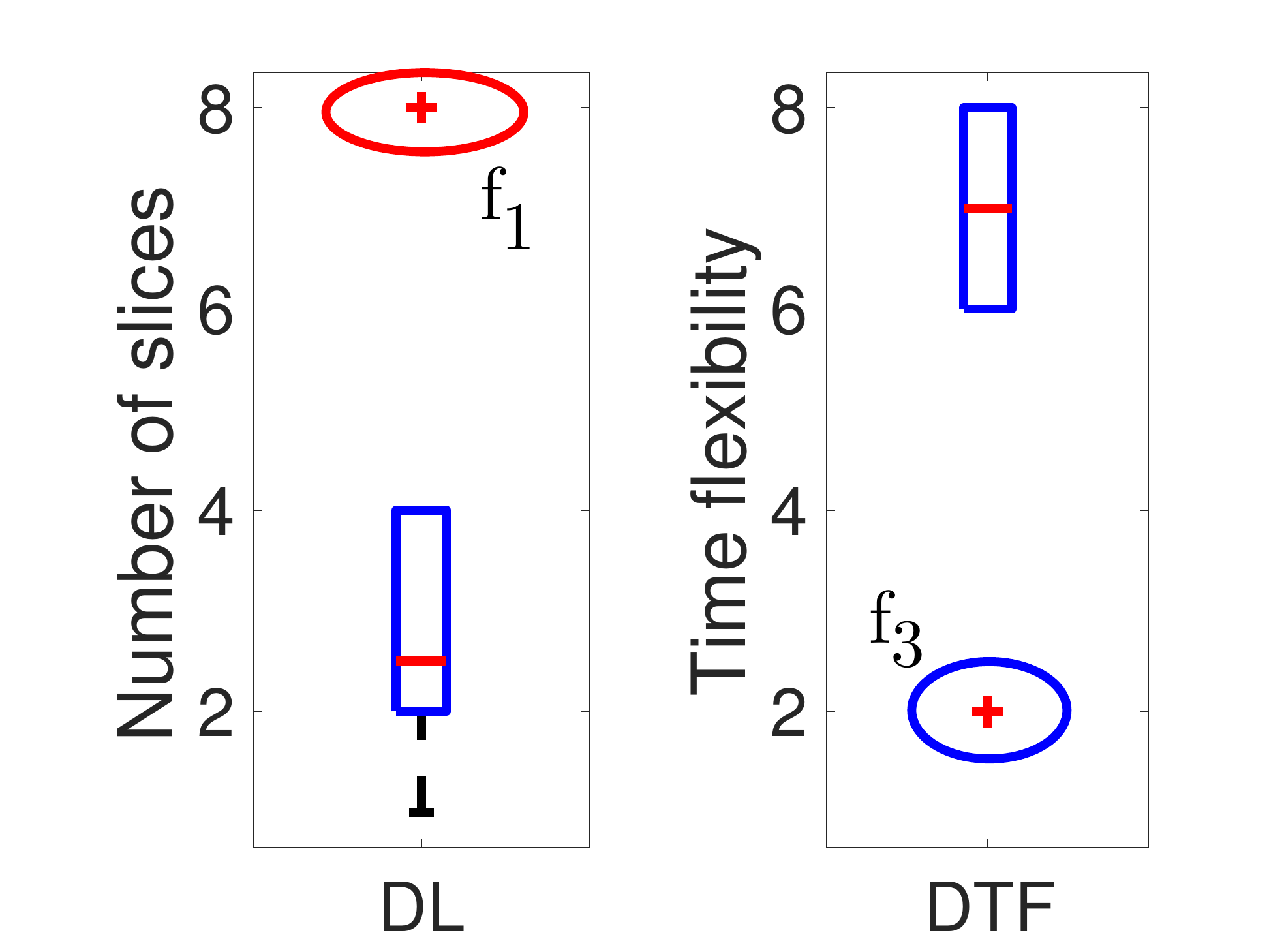}\\
(a) & (b) \hspace{4.8em}  (c)  \\
\end{tabular}
\caption{DL and DTF example,  profile size and time flexibility box plots \label{fig:dldtfex}} 
\end{figure}

\textbf{DP - Initialization phase.}
During the initialization phase, DP divides the initial set $F$  into $2$ subsets.
First, DP computes the upper fence  ($\mathit{uf}$)~\cite{boxplot} of the power profile size of the FOs in $F$ (\refalg{dp}~\reflin{dpuf}).
Then, it stores in $\mathit{PF}$ the FOs that have profile size of at most $\mathit{uf}$ (\reflin{dpal}). 
It selects  as  $f_{ini}$   the most flexible FO  in $\mathit{PF}$ among the ones  with the   longest profile      and removes it from $\mathit{PF}$ (\reflinto{dpselectfnom}{dpreturn}).
For instance, given the  set $F$ in~\reffig{dldtfex}a
(\begin{math}\{f_1,\dots,f_6\end{math}\}), $\mathit{uf}$ is $4$, see~\reffig{dldtfex}b.
DP excludes  $f_1$, which has a very long profile compared to the other FOs (red circle in~\reffig{dldtfex}b), from $F$ and selects   FO $f_6$ as $f_{ini}$. 
FOs with very long profiles  have difficulties satisfying the slice equality and it is likely that they have small time flexibility due to their long profiles (e.g., many charging hours for the EVs).
Thus, they have less potential alignments to further satisfy the flexible order requirements.
Then, DP continues aggregation  with the processing and examination phase using    $\mathit{PF}$, i.e., \begin{math}F\setminus \{f_{ini}\cup f_1\}\end{math}.

 
\textbf{DTF - Initialization phase.}
DTF takes into account the time flexibility distribution of the initial set $F$ and      excludes FOs with low time flexibility compared to the initial set.
It computes the lower fence of time flexibility distribution of  $F$    and sets the time flexibility threshold ($\mathit{tft}$) equal to the lower fence~\cite{boxplot} (\refalg{dtf}~\reflin{dtflf}).
It splits    $F$ based on the lower fence of the  time flexibility distribution in the set.
It stores the FOs with time flexibility at least $\mathit{tft}$ in $\mathit{PF}$ (\reflin{dtfsplit1}).
DTF then   selects $f_{ini}$ from $\mathit{PF}$ (\reflin{dtfini}).
As a result, the algorithm excludes the FOs that have very small time flexibility.
For instance, given the set $F$ in~\reffig{dldtfex}a, $\mathit{tft}$ equals   $6$,~\reffig{dldtfex}c.
Thus, DTF excludes  $f_3$, which has very low time flexibility compared to the other \fo{}s in the set, from $F$,  see the blue circle in~\reffig{dldtfex}c. 
DTF then sets $\mathit{tft}$ to $6$, selects   FO $f_1$ as $f_{ini}$, and continues aggregation with $\mathit{PF}$, i.e.,  \begin{math}F\setminus \{f_{ini}\cup f_3\}\end{math}.
FOs with small time flexibility have lower probability to contribute in aggregation due to the low number of alignments that they have.
Moreover,  by setting $\mathit{tft}$  equal to the lower fence, DTF reduces the number of   examined alignments and consequently the complexity of the algorithm.
Thus, AFOs with greater time flexibility are more likely to be produced.   

\begin{algorithm}[tb]
\begin{algorithmic}[1]
\Function{Initialize}{$F$}
\State $\mathit{tft}\leftarrow$LowerFenceTimeFlexibility($F$)\label{lin:dtflf}
\State $\mathit{PF}\leftarrow$FOsWithTimeFlexibilityAtLeast$\mathit{tft}$($F,\mathit{tft}$) \label{lin:dtfsplit1} 
\State $f_{ini}\leftarrow$SelectTheMostFlexibleFOAmongLongest($\mathit{PF}$)\label{lin:dtfini}
\State\Return $ \mathit{PF}\setminus f_{ini}, F\setminus\mathit{PF},f_{ini},\mathit{tft}$
\EndFunction
\end{algorithmic}
\caption{Initialization phase - Dynamic Time Flexibility}\label{alg:dtf}
\end{algorithm} 

\textbf{Processing  phase.}
In the processing phase,   \ma{}   
    examines all the potential binary aggregations between  $f_{ini}$ and the   FOs in       $\mathit{PF}$ defined in the initialization phase.
The FOs   are examined   in   descending order according to their time flexibility. 
FOs with high time flexibility have more potential to participate in an aggregation that fulfills the flexible order requirements because of  high number of alignments. 

\begin{algorithm}[tb]
\begin{algorithmic}[1]
\Function{Process}{$\mathit{PF},\mathit{AF}, f_{ini},\mathit{tft},\mathit{ppt},\mathit{spt},e$}
\State  $\mathit{PF}_{tmp} \leftarrow \emptyset$, $\mathit{f}_{a} \leftarrow$ null 
\ForAll {   $f\in \mathit{PF}$ }\label{lin:ppfirstfor}
	\State$f_{cand}\leftarrow$ null, $bestCV\leftarrow \infty$   	 
 	\ForAll {alignment $al$ of $\{f_{ini}, f\}$ }\label{lin:ppsecondfor}
		\State $f_x\leftarrow$BinaryAggregation($f_{ini},f,al,\mathit{tft},\mathit{ppt}$)
		\If{RMSE($f_x, \mathit{spt}$)$<$RMSE($f_{ini}, \mathit{spt}$)}\label{lin:pprmse}	  
 			\If{CV($f_x$)$<$$ bestCV$}\label{lin:ppcv}
				\State $bestCV\leftarrow$CV($f_x$)\label{lin:ppcv},  $f_{cand}\leftarrow f_{x}$
			\EndIf
		\EndIf\label{lin:ppconend}
	\EndFor
	\If{ $f_{cand} \neq$ null} 
			\State  $\mathit{PF}_{tmp} \leftarrow \mathit{PF}_{tmp} \cup f$\label{lin:pptmp},   $f_{ini} \leftarrow f_{cand} $
	\EndIf
	\If{ $\forall   s \in P( f_{ini} ), spt -e < s.p < spt +e$} \label{lin:tlcond}
			\State  $f_a  \leftarrow  f_{ini}$\label{lin:florder}	
			\State  $\mathit{PF}  \leftarrow  \mathit{PF} \setminus \mathit{PF}_{tmp}$  \label{lin:ppdel}		
 			\State  $\mathit{PF}_{tmp} \leftarrow      \emptyset$, $\mathit{spt}  \leftarrow  \mathit{spt} $$+$$100$  \label{lin:tlincrease}		
	\EndIf
\EndFor
\State \Return  $ \mathit{PF}$, $\mathit{AF} \cup f_a$\label{lin:ppreturn}	 
 	\EndFunction
\end{algorithmic}
\caption{Processing phase}\label{alg:pp}
\end{algorithm}

\begin{algorithm}[tb]
\begin{algorithmic}[1]
\Function{Examine}{$\mathit{PF,UF, AF}, continue$}
\If{$\mathit{PF\cup UF}$$=$$\emptyset$ \textbf{or}   ($\mathit{|AF|}$$\geq$$5$ \textbf{and} \par\hskip\algorithmicindent 
totalEnergy($\mathit{PF}$$\cup$$\mathit{UF}$)$<$Energy5\textsuperscript{th}AFO($\mathit{AF}$))} \label{lin:exfirstif}
\State 	 $continue \leftarrow false$\label{lin:setcontinuefalse}
\EndIf
\State\Return     $\mathit{PF\cup UF}$, $ continue$\label{lin:exreturn}
 	\EndFunction
\end{algorithmic}
\caption{Examination phase}\label{alg:ex}
\end{algorithm} 

\ma{} examines, through   the potential alignments, all the   binary aggregations that fulfill the time flexibility  $\mathit{tft}$ and the power profile thresholds $\mathit{ppt}$  (\refalg{pp},~\reflinto{ppfirstfor}{ppsecondfor}).
Among the AFOs that reduce the root mean square error (RMSE) between $f_{ini}$ and the slice power threshold $\mathit{spt}$, it chooses the one with the minimum coefficient of variation (CV) (\reflinto{pprmse}{ppconend}). 
By promoting the reduction of RMSE, the produced AFO $f_{cand}$ has a power profile closer to $\mathit{spt}$.
In particular, the use of RMSE during aggregation prevents the  increase of profile length of the potential AFO and contributes to the production of slices with values closer to $\mathit{spt}$. 
Consequently,  alignments   that lead to power profiles that time-wise overlap each other are preferred for aggregation.
Moreover, because the slices of an AFO might have power deviations, the second condition of CV (\reflin{ppcv}) is used.
A low CV  of $f_{cand}$ contributes to the elimination of power profile deviations and to the production of  AFOs with slice power amounts closer to each other.
For instance, given the FOs  in~\reffig{alignments}  and  $\mathit{spt}$ equal to $3$, the RMSE between the slices of AFO $f_{12}$ and $\mathit{spt}$   is equal to $1$ and lower than the RMSE between the longest FO $f_{123}$  and $\mathit{spt}$, which is $1.8$.
Similarly,    $f_{12}$ and $f_{123}$ have CV equal to $0$ and $0.4$, respectively, with $f_{12}$ having no power fluctuations.
Thus, the reduction of RMSE and CV lead to AFOs that fulfill the flexible order energy requirements. 

\begin{figure*}[tb]
\begin{center}
\begin{tabular}{cccc}
\hspace{-1em}
\includegraphics[width=0.24\textwidth]{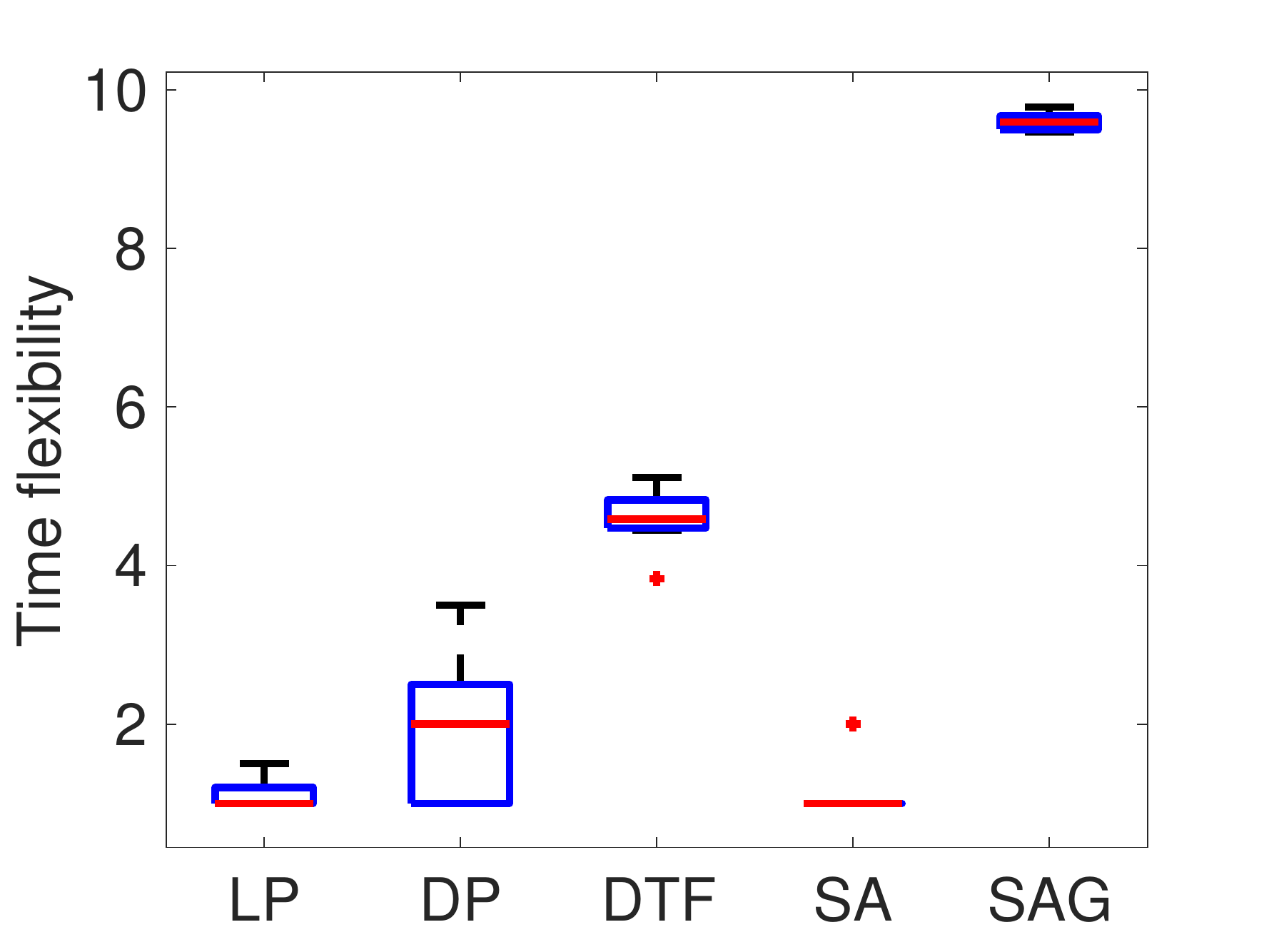}&	\includegraphics[width=0.24\textwidth]{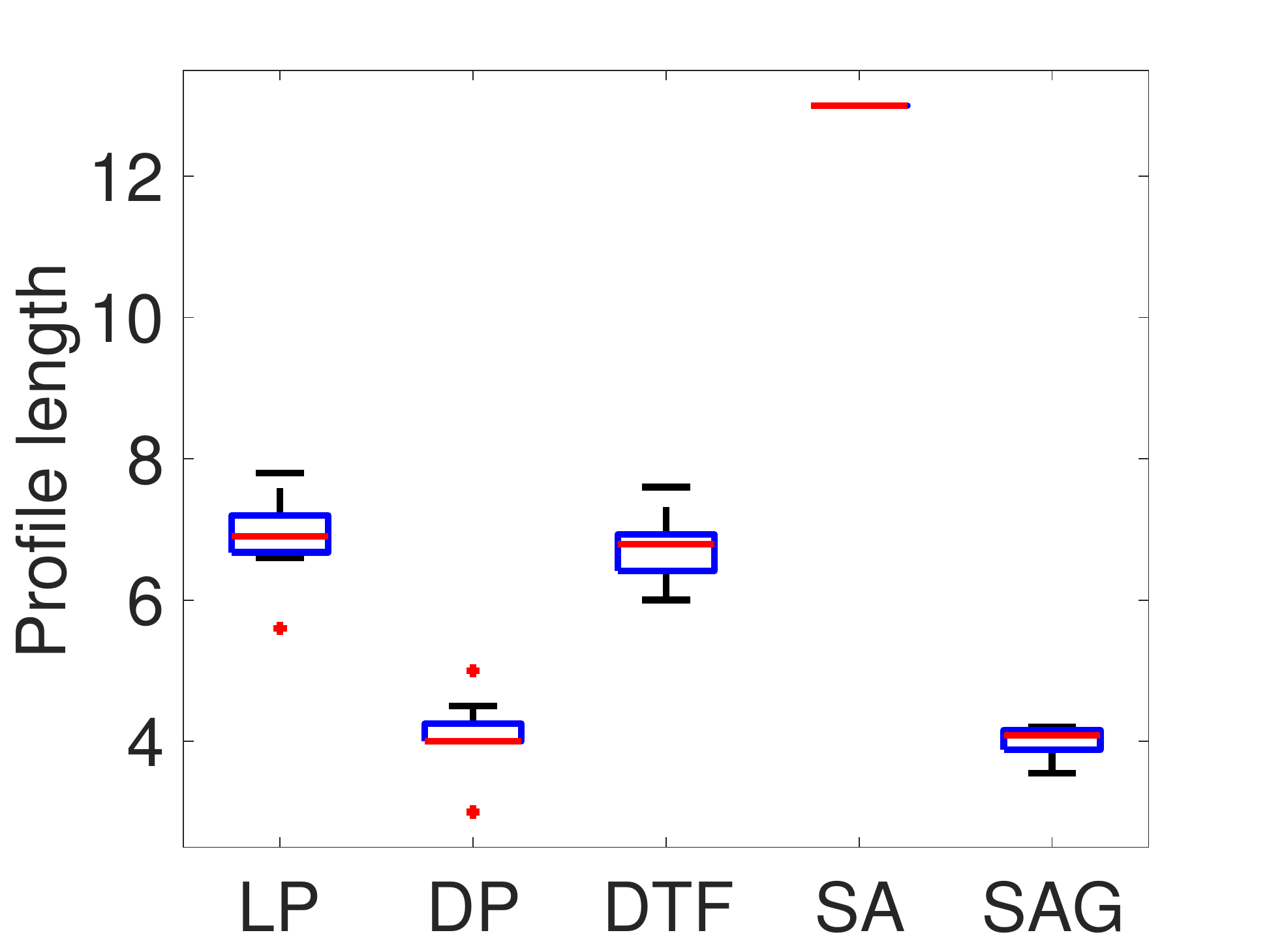}&	
\includegraphics[width=0.24\textwidth]{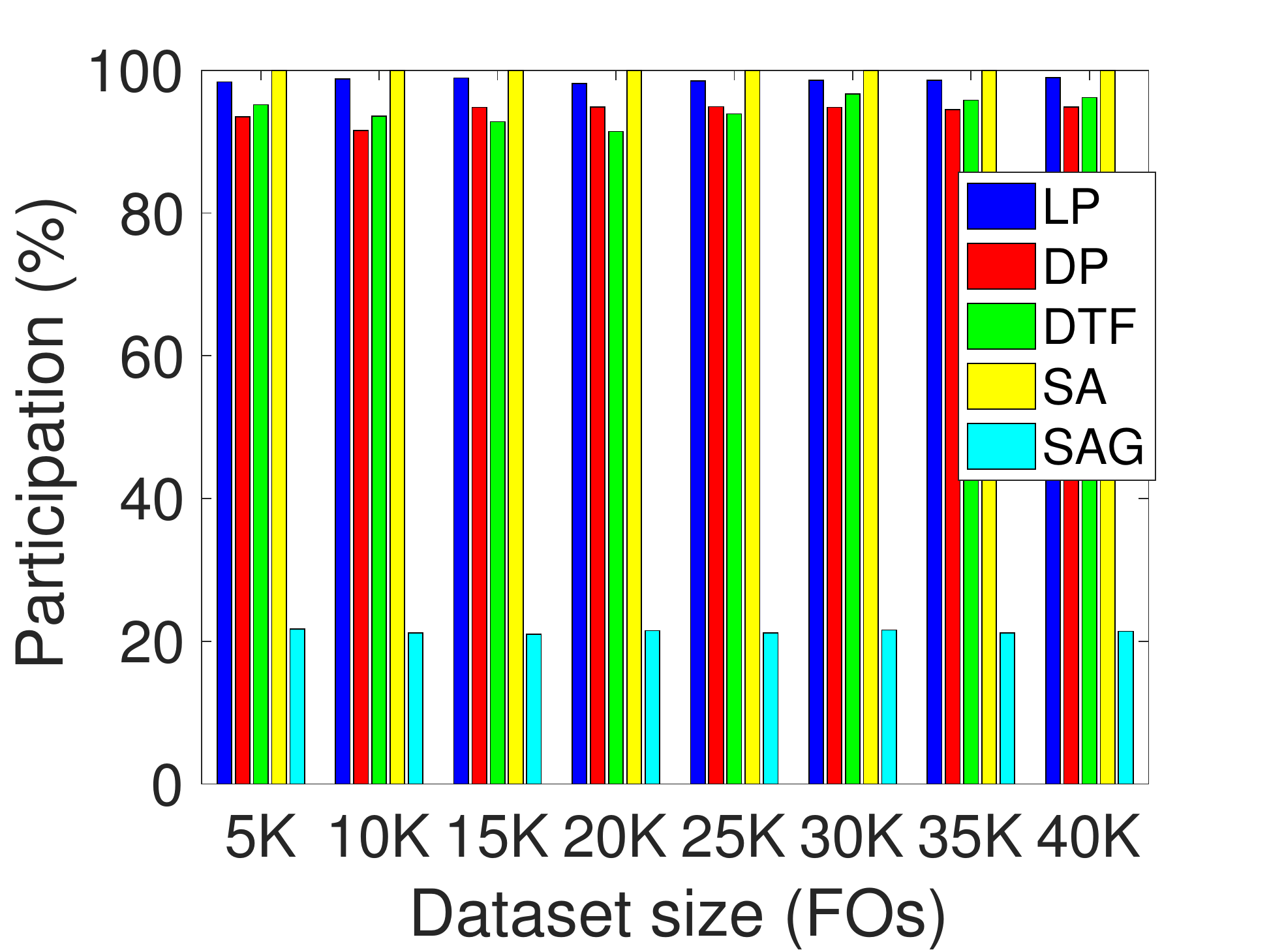} &
\includegraphics[width=0.24\textwidth]{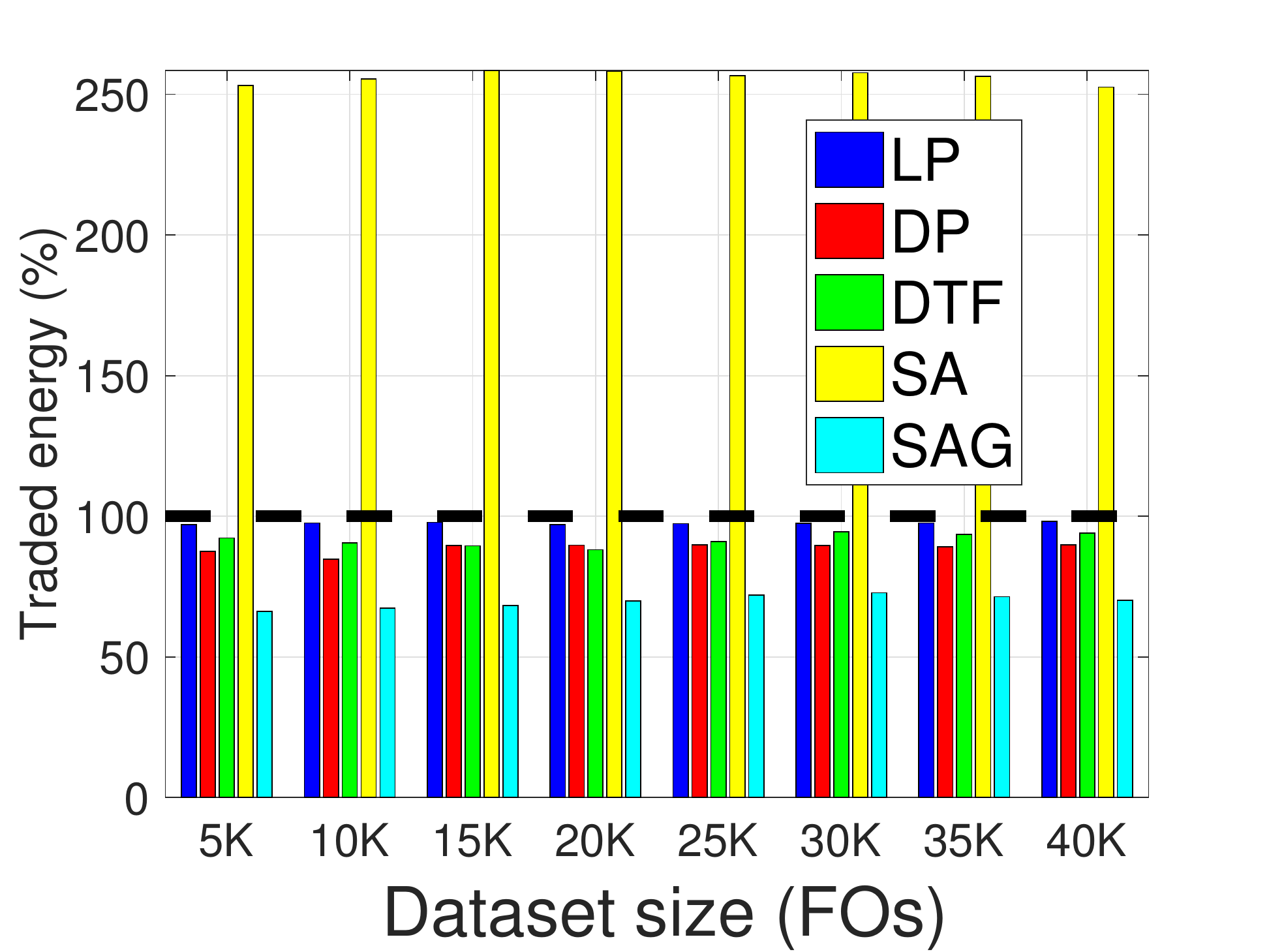} \\
(a) Average time flexibility	&	 
(b) Average profile length&	 
(c) Participation in aggregation &
(d) Traded energy\\
       \end{tabular}
  \caption{Average time flexibility, average profile length, participation of FOs,  and traded energy}
	\label{fig:expFig1}  
  \end{center}
\end{figure*} 

When an AFO with power amounts around $\mathit{spt}$ is produced, 
an  $e$ kW deviation per slice   is permitted (\refalg{pp}~\reflin{tlcond}).
At that point, an AFO $f_a$ that fulfills the flexible order criteria is produced (\reflin{florder}).
The FOs that participate in aggregation are temporally stored (\reflin{pptmp}) and when an AFO $f_a$ is produced, they are removed from $\mathit{PF}$ (\reflin{ppdel}).
Then, $\mathit{spt}$ is increased by 100 (\reflin{tlincrease}) so that AFOs with larger energy are produced during the following aggregation.
As a result, the processing phase produces an AFO that captures large amounts of energy and fulfills the time flexibility and power amount requirements of a flexible order.
When all the FOs in $\mathit{PF}$ are processed, \ma{} 
returns both $\mathit{PF}$ and 
the output set $\mathit{AF}$ with the aggregated FO $f_a$ (\reflin{ppreturn}).
 
\textbf{Examination  phase.}
During the examination phase, \ma{} first examines if there are any FOs in either $\mathit{PF}$ or $\mathit{UF}$ to further continue  aggregation (\refalg{ex}~\reflin{exfirstif}).
In case,   
the total  energy of the remaining FOs is larger than the $5$th in descending size  energy AFO,
\ma{} continues using the remaining FOs (\reflin{exreturn}). 
Otherwise,~\ma{} does not continue the execution  (\reflin{setcontinuefalse}). 
As a result, the algorithm ensures that the remaining FOs cannot produce an AFO with energy greater than one of the $5$ produced AFOs.
Since   the $5$ AFOs with the most energy will be transformed into flexible orders, the algorithm terminates (\refalg{ha}~\reflin{hareturn}).

\section{Experimental Evaluation}
\label{sec:exp}
\subsection{Experimental setup}

We consider a BRP managing a portfolio of EVs represented by \fo{}s.
The BRP utilizes our proposed aggregation algorithms to produce AFOs that respect the flexible order requirements.
The BRP transforms the $5$ AFOs
which capture the highest amount of  energy  into flexible orders and trades them in Elspot.
In order to  examine the scalability of our proposed algorithms,
we create $8$ differently-sized \fo{} datasets, from $5$K to $40$K \fo{}s (multiples of $5$K), 
with characteristics based on the probability distributions suggested in~\cite{7098444}.
Moreover, we consider that all EVs use the charging option described in~\refsec{EVmodel} and need to be fully charged. 
Thus, the   initial SOC of all EVs is within [$20$\%, $85$\%], while they must be charged  up to $90$\%.
Details about the characteristics of the datasets are in~\reftab{characteristics}.

We compare our techniques with two baseline aggregation techniques~\cite{TKDEEManolis}.
We use Start-Alignment (SA) aggregation, see~\refsec{SA} and Start-Alignment Grouping (SAG) aggregation.
SAG groups together FOs that have both the same earliest start charging time and the same time flexibility and then applies SA on each group.
As a result, it produces one AFO per group. 
We evaluate our techniques in terms of output size (\#AFOs), participation of FOs in aggregation, percentage of energy traded in the market, running time, and both time flexibility and profile length of AFOs.

\subsection{Market-based aggregation results}

\textbf{Output size.}
SA always produces one AFO whereas SAG produces more than $100$ AFOs in all cases. 
Both LP and DP produce less than or equal to $5$ AFOs in all cases.
DTF produces more than $5$ AFOs in  $75$\% of the cases as
the energy threshold is activated  in a later step compared to the other techniques due to the division of the processed set.

\begin{figure*}[tb]
\begin{center}
\begin{tabular}{cccc}
\hspace{-1em}\includegraphics[width=0.24\textwidth]{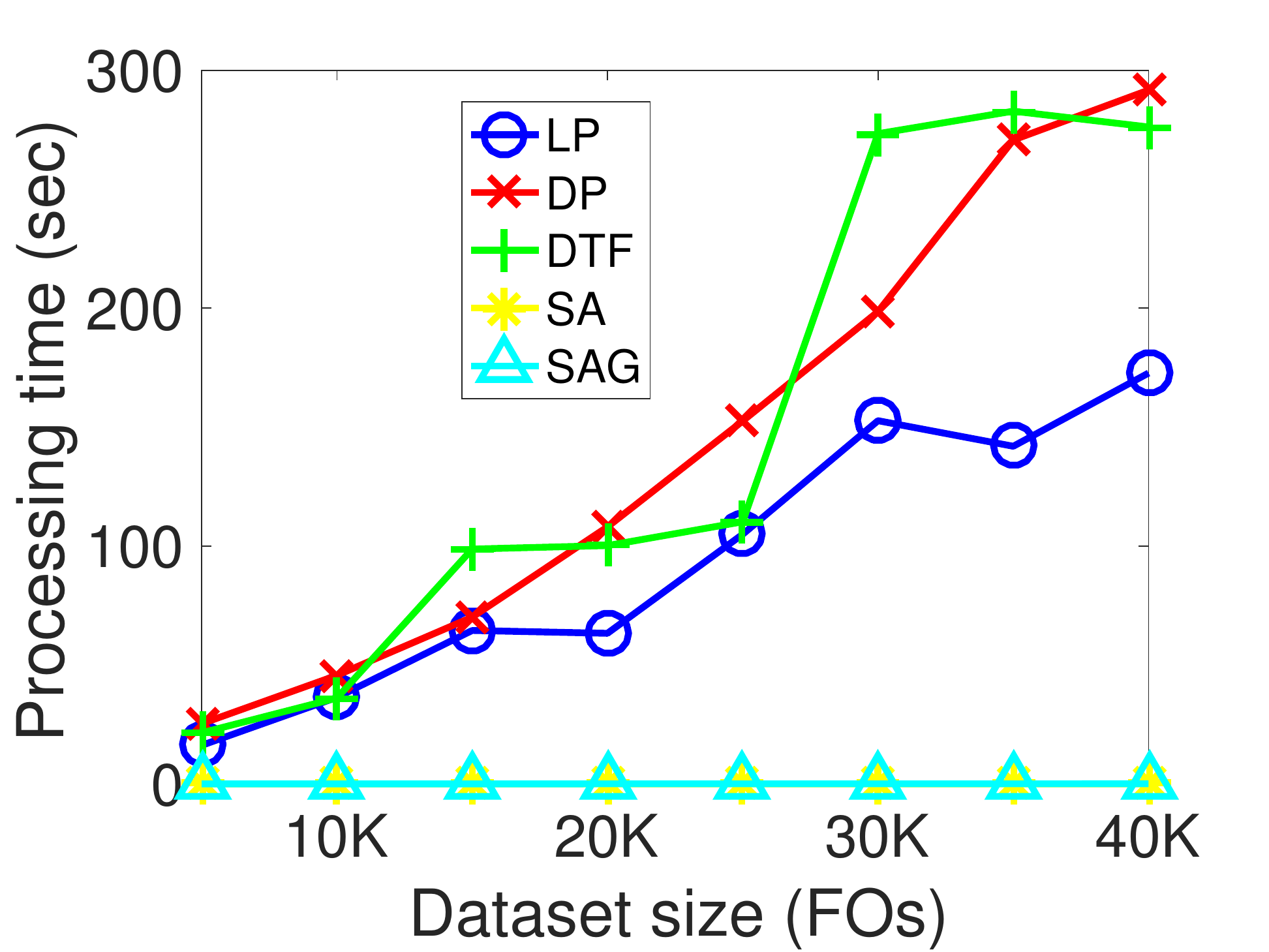}&
\includegraphics[width=0.24\textwidth]{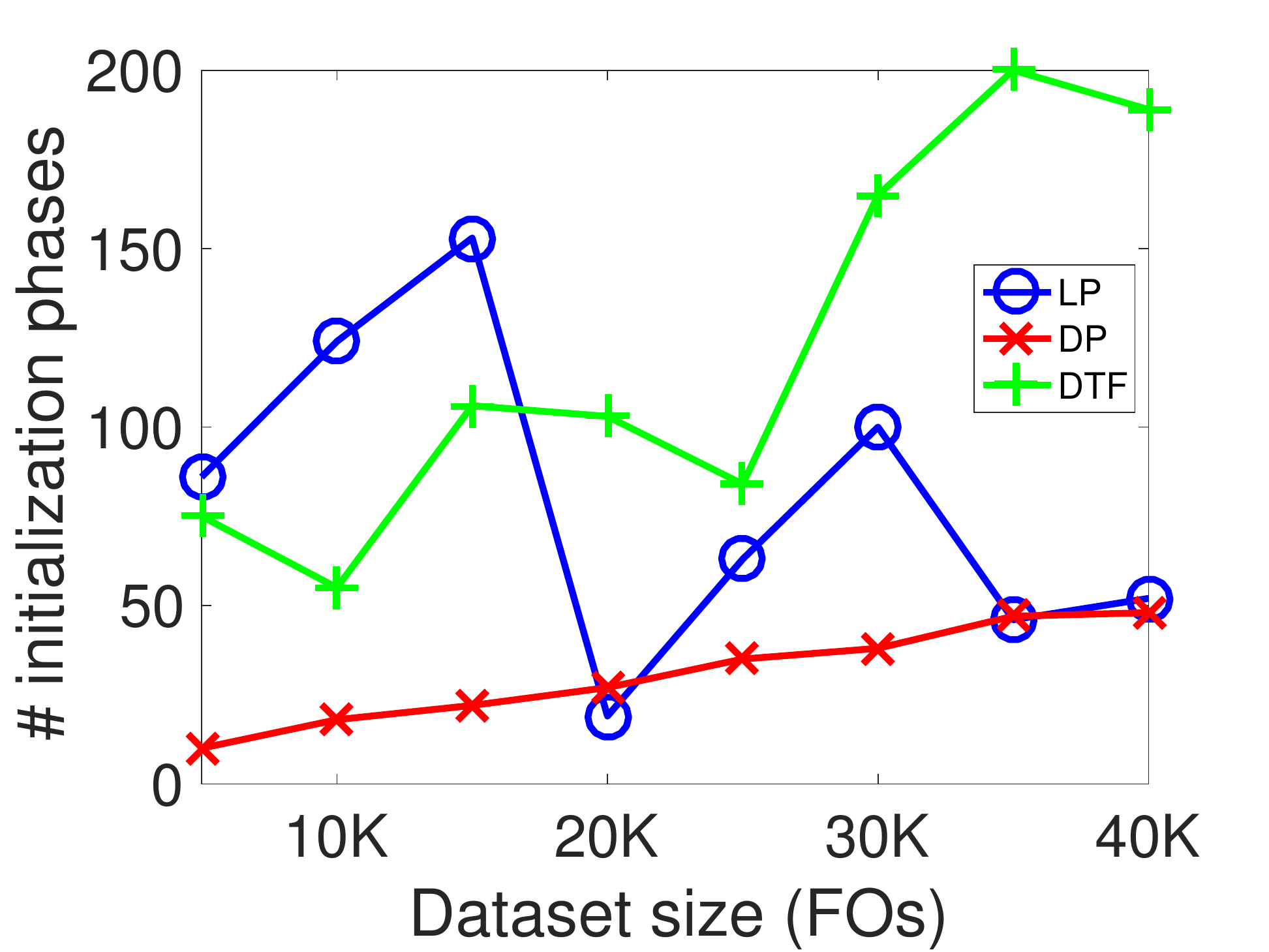} &  	\includegraphics[width=0.24\textwidth]{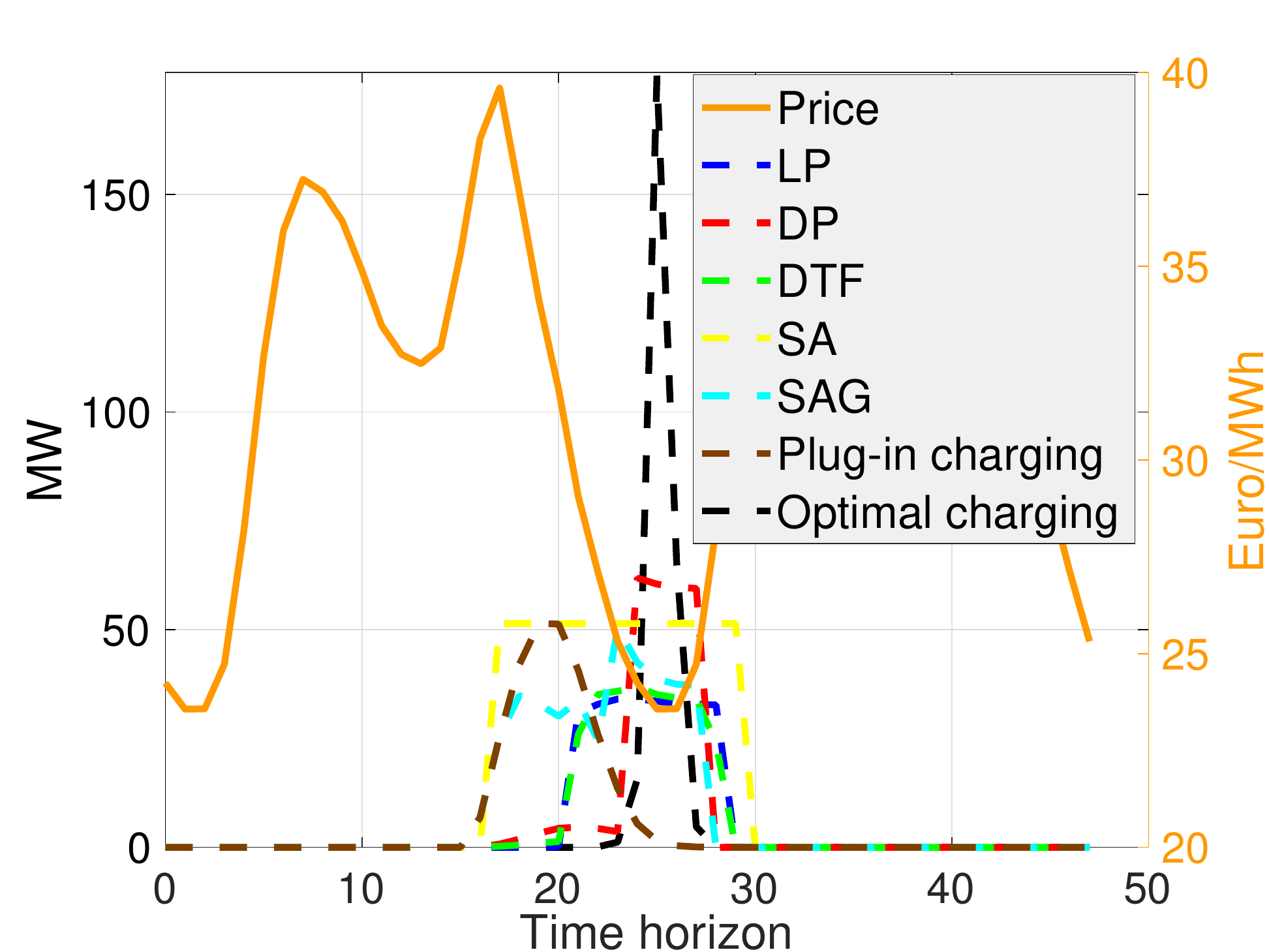}&
\includegraphics[width=0.23\textwidth]{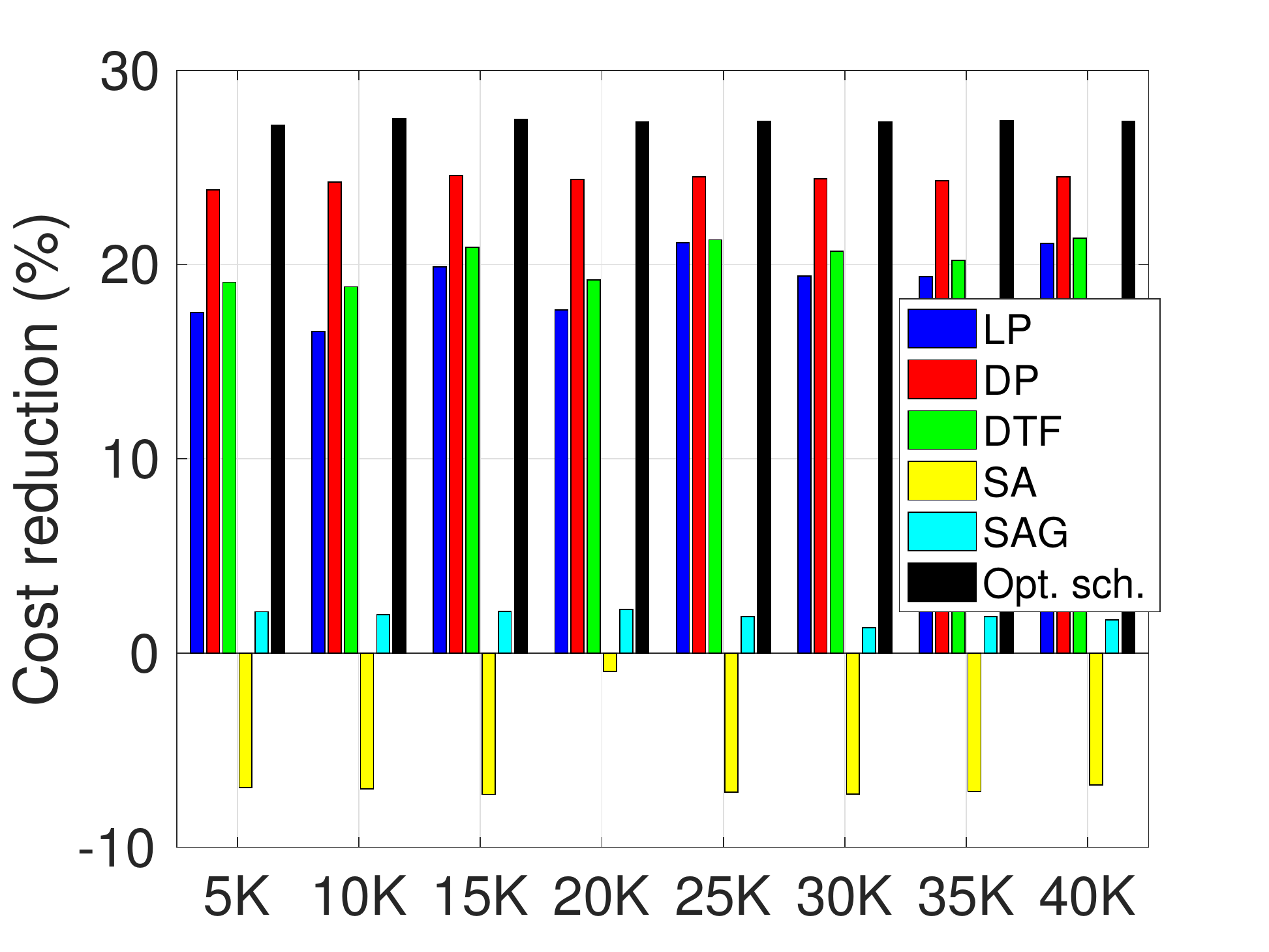}\\	
(a) Processing time&	
(b) \# of initialization phases&	
(c) Charging time vs pricing	&
(d) Cost reduction\\ 
       \end{tabular}
  \caption{Processing time, number of initialization phases, charging times and pricing for 40K
  FOs  dataset, and cost reduction for all datasets}
	\label{fig:expFig2}  
  \end{center}
\end{figure*}

\textbf{Time flexibility and profile length.}
Regarding the   baseline techniques, 
SA produces   long AFOs with very low time flexibility  as it aggregates all FOs into one.
On the contrary, SAG  produces  short and time flexible AFOs due to the grouping phase it applies,  see~\reffig{expFig1}a, b.
LP uses as initial FO ($f_{\mathit{ini}}$) the longest FO of the dataset. 
Usually, such an FO has low time flexibility and so do the produced AFOs.
Due to the long profile of $f_{\mathit{ini}}$, LP might utilize all the time flexibility of the remaining FOs to produce an AFO that reduces the distance to the power profile threshold ($\mathit{ppt}$).
Consequently, LP produces long AFOs with very low time flexibility, see~\reffig{expFig1}a, b.
The AFOs produced by DP are more flexible than the ones from LP since DP applies a dynamic profile size approach   and excludes from aggregation very long FOs.
As a result, FOs with similar profiles are aggregated together and less time flexibility is required to find a proper alignment that   minimizes the distance to $\mathit{ppt}$.
Consequently, AFOs with less slices compared to LP are produced, see~\reffig{expFig1}b.
Finally, DTF produces the most flexible AFOs among our proposed techniques.
We see in~\reffig{expFig1}a that the average time flexibility of the produced AFOs is  greater than $4$ in all datasets.
DTF achieves it by utilizing the time flexibility threshold.
However, DTF produces long AFOs, similar to LP, because it also selects as $f_{\mathit{ini}}$ the longest AFO of the processed set, see~\reffig{expFig1}b.

\begin{table}[tb] 
\centering
\begin{tabular}{ |c ||c |c |c |c| c|}
  \hline                       
      &    Distr.&   Mean &St. dev &  Min  &    Max  \\  \hline \hline       
Battery capacity   (kWh)&    UD$*$ &   $23$& 4 & 16&  30 \\  \hline                 
Arrival time&    TGD$*$  &   $19$$:$$00$&  2h & $16$$:$$00$ & $1$$:$$00$       \\  \hline                 
Departure time&     TGD$*$ &  $7$$:$$00$&   2h & $5$$:$$00$ &  $12$$:$$00$   \\  \hline             
Initial Battery SOE (\%)&    TGD$*$  &  75&$25$& $20$&   $85$  \\  \hline                                       
\end{tabular} 
 $*$ UD: uniform distribution, TGD: truncated Gaussian distribution
\caption{EV data probability distribution\label{tab:characteristics}}
\end{table} 

\textbf{Participation and traded energy.} 
In order to quantify the participation of  FOs in aggregation, 
we take into account only the FOs that participate in the aggregation of the $5$ (or less) largest in energy AFOs, 
i.e., the AFOs that are transformed into flexible orders.
Similarly, we compute the traded energy by taking into account only the energy captured by the AFOs that are transformed into flexible orders.


SA aggregates all FOs into one AFO and thus participation in aggregation is $100$\%, see~\reffig{expFig1}c.
The slices of the AFO have very high power differences and since a flexible order requires a flat   power profile, 
the power of the highest slice  is considered for the whole profile of the AFO.
As a result, on average, $2.5$ times the energy captured by that AFO is traded, see~\reffig{expFig1}d where $100$\% is the energy needed for all the FOs.
On the contrary, SAG produces too many AFOs and since only the $5$ largest are traded, 
we see a very low participation percentage and
the lowest percentage of traded energy among the techniques ($69.7$\% on average).
In general, the longest AFOs capture  more energy as they have more slices and more FOs participate in their aggregation. 
Thus, LP, which produces the longest AFOs, obtains both the highest participation percentage ($98.6$\%) and   traded energy   percentage ($97.5$\%) in all the cases among our proposed techniques, see in~\reffig{expFig1}c, d.
DTF follows with an average participation value of   $94.4\%$ and $91.7\%$ percentage of traded energy.
DP has the lowest percentage in both participation and traded energy, 
$94.2$\% and $88.8$\% on average respectively.
The reason is that DP excludes   very long FOs, which usually capture large energy, from aggregation.
 
\textbf{Processing time.} 
Both SA and SAG are  fast techniques with processing times below one second as they examine a  very small solution space and do not consider the market requirements.
LP is the fastest among all our proposed techniques since it efficiently activates the energy threshold, see~\reffig{expFig2}a.
The processing time of DP follows a close to linear growth rate.
DTF has an increasing trend for processing time, but it shows similar processing times for datasets with different sizes, e.g., for datasets with $30$K and $35$K FOs.
The reason is that the processing time is highly driven by the number of initialization phases.
The size of the dataset might   increase, but the new added FOs might lead to less initialization phases and therefore to less aggregation comparisons.
That is why we also notice that both processing time and  number of initialization phases follow similar patterns. 
Whenever the number of initialization phases is increased compared to the previous dataset, processing time also increases.
For instance, we see in~\reffig{expFig2}b that when the size of the dataset is increased from $15$K to $20$K for both LP and DTF, the number of initialization phases is reduced.
As a result, the processing time is similar for both the datasets and slightly increases for the $25$K.
Eventually, when the size of the dataset is further increased, it becomes more difficult for DTF to fulfill the market requirements and thus   both the initialization phases and the processing time are highly increased.

\subsection{Financial evaluation} 

Since the overall goal of a BRP is to trade the AFOs   in the market using flexible orders, we   financially evaluate our aggregation techniques. 
We compare the cost of buying the energy needed to charge the EVs based on plug-in time (traditional approach)
with the cost of charging the EVs by utilizing flexible orders.
Moreover, in order to compare our techniques with the optimal solution,
we   consider a  {\em non-realizable in practice}  scenario where each FO directly participates in the market without aggregation 
and each EV is charged when the charging cost is minimized.
 
Due to the fact that flexibility appears during the night~\cite{6461500}, 
we consider a $48$ hours trading period with a repetition of the $24$h Elspot average prices of $2017$~\cite{Nordpool}, see price curve in~\reffig{expFig2}c.
In the same figure, we illustrate the time and the energy amount
used to charge the $40$K dataset based on our techniques, the two baseline techniques, the plug-in times of the EVs, and     the optimal charging.
We  see that the  charging of the EVs based on the plug-in time occurs when the prices are still high and it does not take advantage of the price drop that occurs in the night of the first $24$ hours.

SA and SAG produce AFOs that do not fulfill the market requirements.
As a result, more energy than     needed has to be traded in the market.
In particular, SA trades    $1.52$ times more energy than   needed to charge the EVs.
Thus, the surplus energy is traded in the regulation market and it results in losses for the BRP, see negative cost reduction  in~\reffig{expFig2}d.
Regarding SAG, the produced AFOs capture  a low   percentage of the energy needed  and they also require extra energy to be traded in order to fulfill the market requirements.
Consequently, the cost reduction due  to the flexible orders trading is eliminated by the losses from the surplus energy trading. 
As a result, we see only $1.1$\% cost reduction on average when SAG is applied. 
On the contrary, the optimal charging option charges all the EVs when the price has the lowest value.
That is why we see a spike in the graph reaching $180$MW after the $24$\textsuperscript{th} hour.

Our proposed aggregation techniques also take advantage of the lowest prices.
LP produces long AFOs which expand over many hours and have low time flexibility. 
That is why we see in~\reffig{expFig2}c that   part of the charging occurs when the prices are high.
DTF produces AFOs that are also long, but they are more flexible than the AFOs produced by LP.
Therefore,  EVs    are charged when prices are a bit lower and DTF achieves a higher cost reduction,~\reffig{expFig2}d.
Finally, DP produces   short and  flexible AFOs.
As a result, it takes advantage of the lowest prices occurring only for a few hours, see~\reffig{expFig2}c.
 
When the energy for the $40$K FOs dataset is purchased based on the plug-in times of the EVs, it costs 8,612 euros.
On the contrary, when LP, with the highest participation, is applied on the $40$K dataset,   39,584  \fo{}s participate in aggregation, see first bar ($98.96\%$) in~\reffig{expFig1}c.
The  39,584  \fo{}s  produce $5$ \afo{}s 
which are further  transformed into flexible orders.
The cost of purchasing the energy needed for the $5$ \afo{}s is computed based on the flexible orders trading and 
it is  6,670  euros, see~\reffig{expFig2}c.
The price also includes the cost ($0.46$ euro) of the imbalances  ($62$kW)  of the flexible orders, see~\refsec{marketFramework}.
The energy needed for the remaining $416$   (40,000$-$39,584) \fo{}s is bought based on their plug-in time and it is $117$ euros.
Thus, the overall energy bought to charge $40$K EVs, when LP is used,   costs   
 6,670$+$126$=$6,796   euros. 
Therefore, LP achieves a $21$\% cost reduction in energy purchase, see LP bar for $40$K dataset in~\reffig{expFig2}d.

We see in~\reffig{expFig2}d that DP achieves on average a 24.4\% cost reduction.
DTF follows with 20.2\% and LP with 19.1\% average cost reduction.
The cost reduction based on the optimal solution is  27.4\%  on average.
Thus,   LP, DTF, and DP   achieve  $69.8$\%, $73.7$\%, and $88.9$\% 
of the optimal cost reduction, respectively.
Notably, the cost reduction that DP achieves {\em only} for the \fo{}s that participate in aggregation   
is on average $98.3$\% of the optimal one.
 
\begin{figure}
\centering
\includegraphics[width=0.4\textwidth]{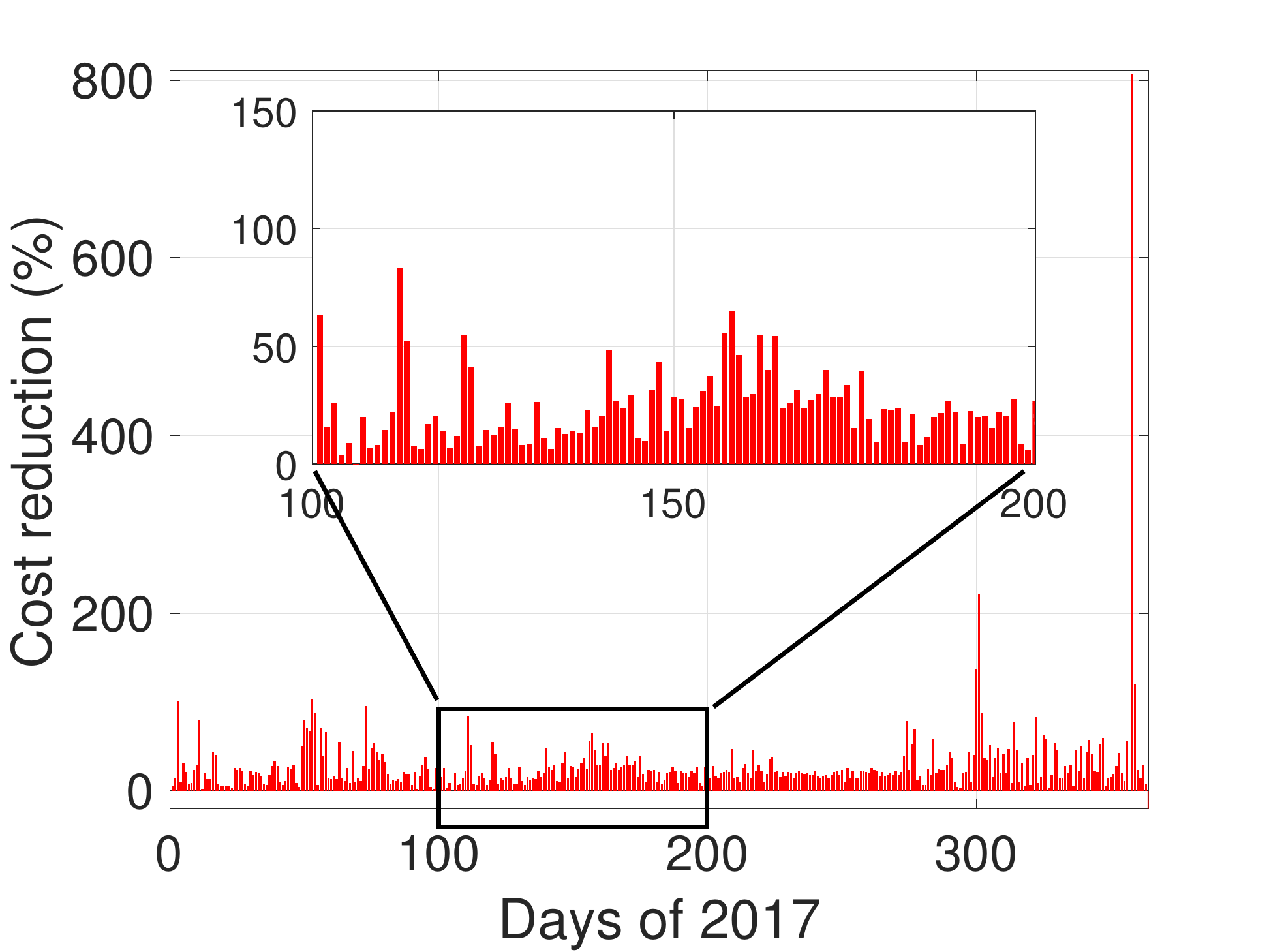}
\caption{Yearly cost reduction based on DP\label{fig:expFig4}  }
\end{figure}

In~\reffig{expFig4}, we illustrate the cost reduction that DP achieves during $2017$. 
We consider $364$ trading periods of  $48$  hours.
The first trading period includes both the  $1$\textsuperscript{st} and the $2$\textsuperscript{nd} day of $2017$. 
The second trading period includes the  $2$\textsuperscript{nd}  and the $3$\textsuperscript{rd}  day of $2017$ and so on.
The average cost reduction is $28$\% and, interestingly, we notice at the end of the year a cost reduction of more than $800$\%.
The reason is  that for several consecutive days, Elspot prices were negative early in the morning and even reached $-$$50$ euros/MWh on the $24$\textsuperscript{th} of December at $2$$:$$0$$0$.

\begin{figure}
\centering
\includegraphics[width=0.4\textwidth]{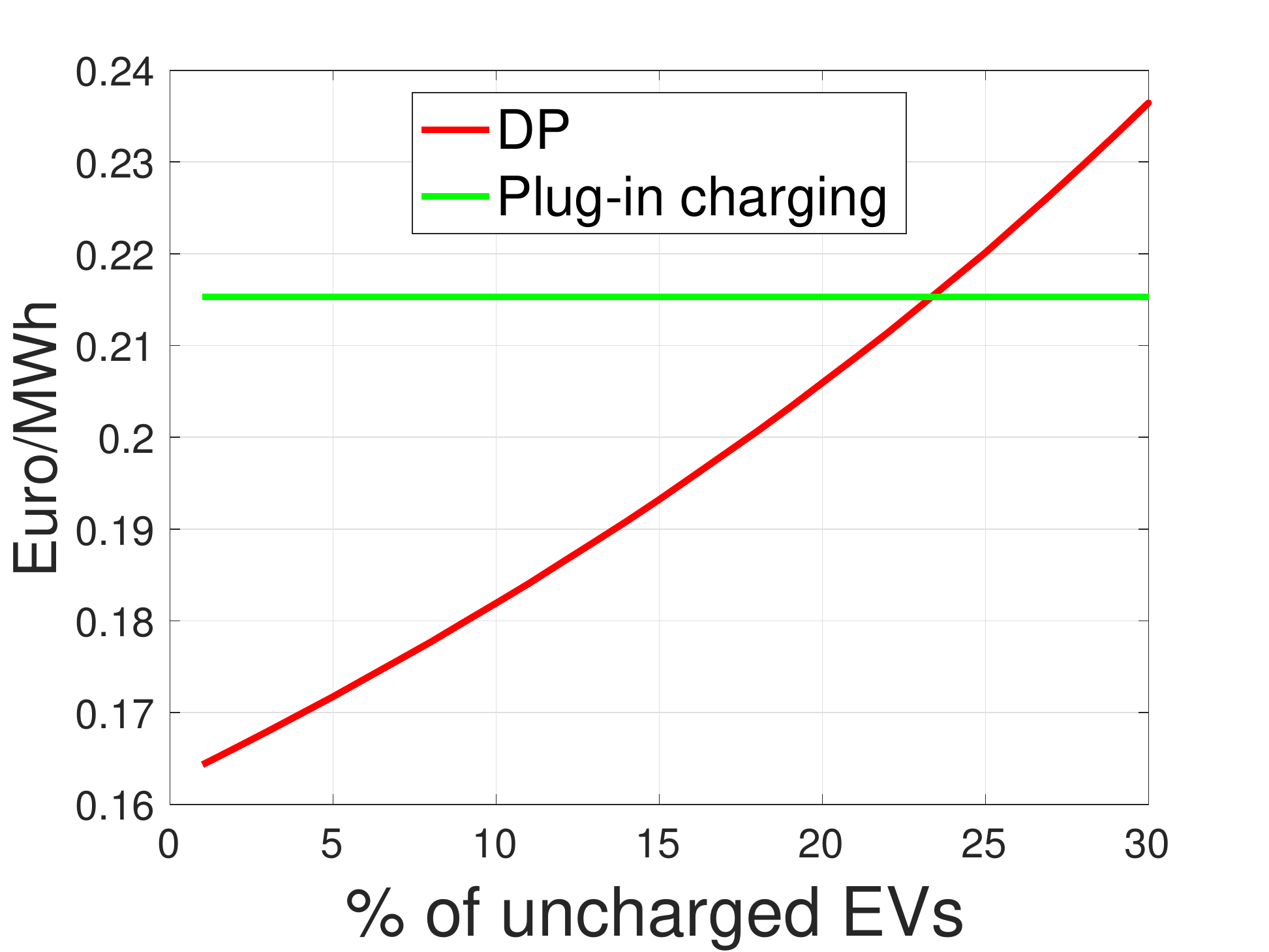}
\caption{Energy price paid by consumers\label{fig:unc}  }
\end{figure}

\textbf{Uncertainty on  FOs forecast.}
The driving behavior can usually be forecasted with very high precision and accuracy.
However, there might be cases where the anticipated flexible load is smaller because  
the EVs are not plugged-in as anticipated.
Thus, we  considered an uncertainty scenario for the  $40$K FOs set according to which 
a percentage of the FOs that participated in aggregation do not need the purchased energy
after the flexible orders derived from DP are placed.
As a result, BRP has to sell back the exceeded purchased amount of energy for a lower price (regulating price).
The difference between the initial cost of  purchasing the energy via flexible orders and the cost of the  exceeded   energy is considered as losses for the BRP.
Moreover, BRP has to distribute (assuming no profit) the purchased energy  and thus the cost to less costumers (EV owners) compared to the initial estimation.
We see in~\reffig{unc}, that the price paid by the  EV owners based on plug-in time is fixed.
On the other hand, the   price for the energy purchased via flexible orders increases as the percentage of the EVs that do not participate in aggregation increases.
The reason is that the cost is higher due to the imbalances and at the same time less consumers use the purchased energy.
We see in~\reffig{unc} that the cost for the consumers via flexible orders is greater than the plug-in charging cost when more than $23$\% of the EVs where imprecisely forecasted.

\textbf{Summary:} By applying our proposed techniques on the aforementioned $365$ trading periods, 
DP achieves the highest cost reduction in $66$\% of the periods and DTF achieves
the highest cost reduction   in the       remaining $34$\% of the periods.
The reason is that the financial impact of the techniques is highly correlated with the pricing curve of the trading period.
Thus, in cases where the price drops for only few hours close to the plug-in charging time, DP is the most suitable technique.
On the other hand, when the price drops for longer periods but much later than the plug-in charging time, DTF achieves a higher reduction than DP. 
\section{Conclusion and Future work}
\label{sec:con}
This paper investigates the market-based   aggregation problem using the FO model to capture  flexible charging loads of EVs.
It proposes $3$ market-based FO aggregation techniques that efficiently aggregate loads from thousand of EVs taking 
into account real market requirements.
Consequently, the techniques produce aggregated FOs that can be   transformed into flexible orders and be traded in the energy market.
The paper financially evaluates the proposed techniques based on real electricity prices and shows that    a   $27$\% cost reduction on energy purchase can be achieved via  flexible orders.  

In our future work,
we will enrich our  techniques considering pricing forecast models and   uncertainty in patterns of driving behavior.
Furthermore, we will investigate more variations of the proposed algorithm and we   prove the theoretical lower bounds       for the their complexity.
Moreover, we will examine  a price-maker market scenario and different market strategies for the BRPs.

\begin{acks}
 This   is work was supported in part by the TotalFlex project funded by the ForskEL program of Energinet.dk and the GoFLEX project funded under the Horizon 2020 program.
\end{acks}
 
\bibliographystyle{ACM-Reference-Format}
\bibliography{bibilio}

\end{document}